\documentclass[journal]{vgtc}                     % final (journal style)
\onlineid{0}

\usepackage{changepage}
\usepackage{graphicx}
\usepackage{xcolor}
\usepackage[most]{tcolorbox}
\usepackage{wrapfig}

\newtcolorbox{fancyquote}{
  enhanced,
  breakable,
  arc=0mm,
  outer arc=0mm,
  colback=white,
  frame hidden,     % <-- removes all faint borders
  borderline west={2pt}{0pt}{gray}, % <-- left gray line only
  left=1em,
  right=0em,
  top=0.0em,
  bottom=0.0em,
  boxsep=0pt
}

\vgtccategory{Research}

\title{Representational Fidelity in Didactic Visualization: Toward a Multidimensional Design Space}

\author{%
  \authororcid{Shehryar Saharan}{0009-0003-9693-8019},
  \authororcid{Michele Oliver}{0000-0003-0776-1235},
  Karen Gordon,
  \authororcid{Ga\"el McGill}{0000-0003-2264-0132}, and
  \authororcid{Jodie Jenkinson}{0000-0002-4066-9732}
}

\authorfooter{
  \item
    Shehryar Saharan and Jodie Jenkinson are with Univ. of Toronto, E-mail: s.saharan@utoronto.ca  |  j.jenkinson@utoronto.ca.
  \item
    Michele Oliver and Karen Gordon are with the Univ. of Guelph, E-mail: moliver@uoguelph.ca  |  kgordon@uoguelph.ca.
  \item
    Ga\"el McGill is with Harvard Medical School, E-mail: mcgill@crystal.harvard.edu.
}

\abstract{%
 Representational fidelity is routinely treated as a single abstract–realistic continuum, a simplification that limits how it is described and compared across research and design contexts. We introduce a \textbf{multidimensional design space of representational fidelity for didactic visualization in science \& engineering}, inductively derived from a 175-item corpus spanning several disciplines, modalities, and instructional aims. The resulting design space specifies five dimensions: Morphological, Dynamic, Cueing, Contextual, and Interactive Fidelity, with seven sub-dimensions. We demonstrate the design space’s descriptive power through successive rounds of expansion and refinement and analyze the corpus to reveal relationships among dimensions and implications for design and research. We further validate the design space through a pilot focus group in which participants applied the dimensions in an open-ended design exercise. Resulting sketches and verbal rationales informed a single-designer applied case study, offering preliminary evidence of the design space's generative potential as a structured aid to design exploration. Together, these contributions lay the groundwork for future research and more intentional design practice. \hyperref[sec:Supplemental]{See Supplemental Materials}.
}

\keywords{Representational fidelity, Didactic visualization, Visual representation, Realism, Abstraction}

\teaser{
  \centering
  \includegraphics[width=\linewidth, alt={A conceptual depiction of the proposed design space of representational fidelity. A designer creates and alters a didactic visualization of blood flow through the heart, adjusting dimensions and subdimensions of representational fidelity, illustrated as continuous, independent dials. The iconography reflects the representational fidelity dimensions that structure the design space presented in the paper. A diagram showing five primary fidelity dimensions—Morphological, Dynamic, Cueing, Contextual, and Interactive—along with their associated sub-dimensions.}]{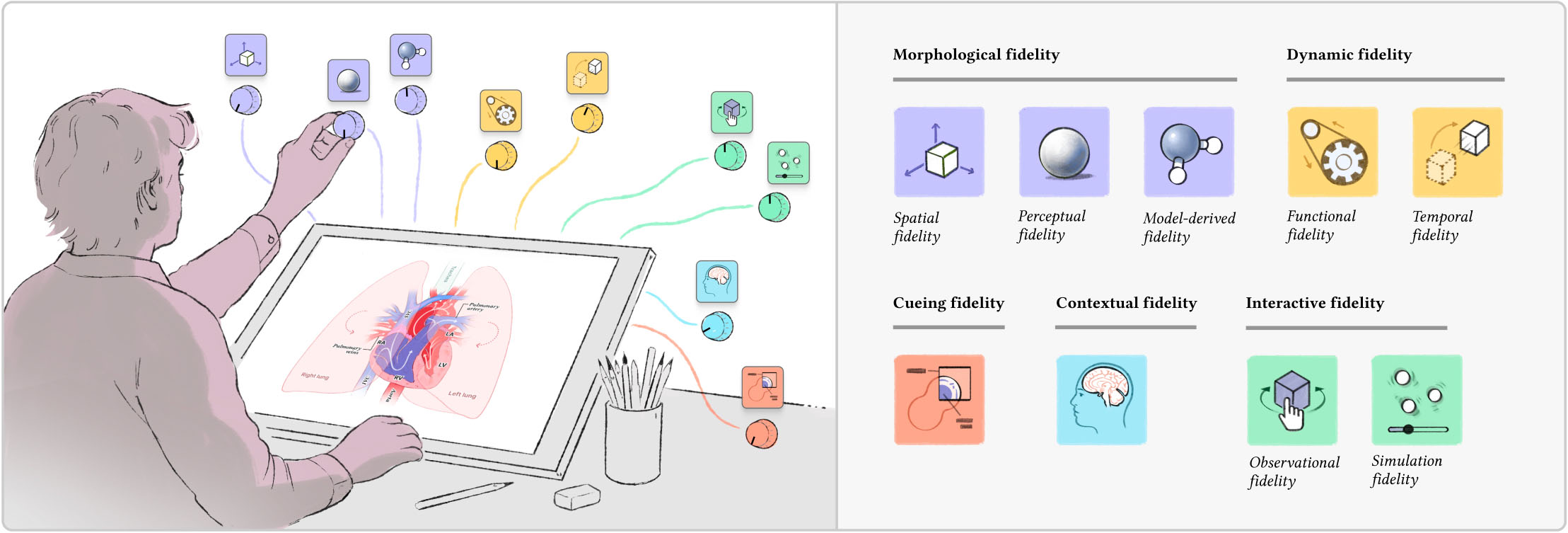}
  \caption{\textbf{Left}: A designer creates a didactic visualization of blood flow through the heart, adjusting dimensions and subdimensions of representational fidelity, illustrated as continuous, independent dials. \textbf{Right}: The proposed design space of representational fidelity, organized across five primary dimensions: Morphological, Dynamic, Cueing, Contextual, and Interactive Fidelity.
  }
  \label{fig:teaser}
}

\graphicspath{{figs/}{figures/}{pictures/}{images/}{./}} % where to search for the images

\usepackage{tabu}                      % only used for the table example
\usepackage{booktabs}                  % only used for the table example
\usepackage{lipsum}                    % used to generate placeholder text
\usepackage{mwe}                       % used to generate placeholder figures
\usepackage{ccicons}                   % package to be able to use icons from creative commons

\usepackage{mathptmx}                  % use matching math font

\begin{document}

%%%%%%%%%%%%%%%%%%%%%%%%%%%%%%%%%%%%%%%%%%%%%%%%%%%%%%%%%%%%%%%%
%%%%%%%%%%%%%%%%%%%%%% START OF THE PAPER %%%%%%%%%%%%%%%%%%%%%%
%%%%%%%%%%%%%%%%%%%%%%%%%%%%%%%%%%%%%%%%%%%%%%%%%%%%%%%%%%%%%%%%

\firstsection{Introduction}

\label{sec:intro2}
\maketitle
Decisions about visual representational fidelity in didactic visualizations (graphical representations designed to educate, explain, or convey instructional concepts) constitute some of the most consequential choices a designer makes: \textit{should a visualization be more ``realistic,'' or strip away detail to highlight the essential structure? Should motion be animated, or staged to scaffold understanding? How much context supports comprehension before it becomes distracting?} Designers of didactic visualizations have long employed a wide range of representational techniques to address these challenges (e.g., moderating complexity \cite{JJComplexity}, experimenting with rendering style \cite{Guerra03042025}, employing annotation strategies \cite{Jian_2014}, etc.) to communicate complex scientific and engineering concepts (that are often multi-scale, dynamic, partially or entirely imperceptible, probabilistic, model-mediated, etc.). Despite the ubiquity of these practices, we lack a shared and systematic way to describe, compare, and reason about representational fidelity for didactic visual media across science and engineering.

\begin{figure*}[t]
  \centering
  \includegraphics[width=\textwidth, alt={Examples of didactic visualizations varying along multiple fidelity dimensions. Four visualizations of the heart and lungs are shown side by side, labeled A–D, ranging from more abstract to more realistic appearances. Each example differs simultaneously in spatial structure, perceptual detail, contextual information, functional depiction, and cueing, illustrating that changes in “fidelity” typically span multiple dimensions rather than a single continuum, especially in experimental visual stimuli}]{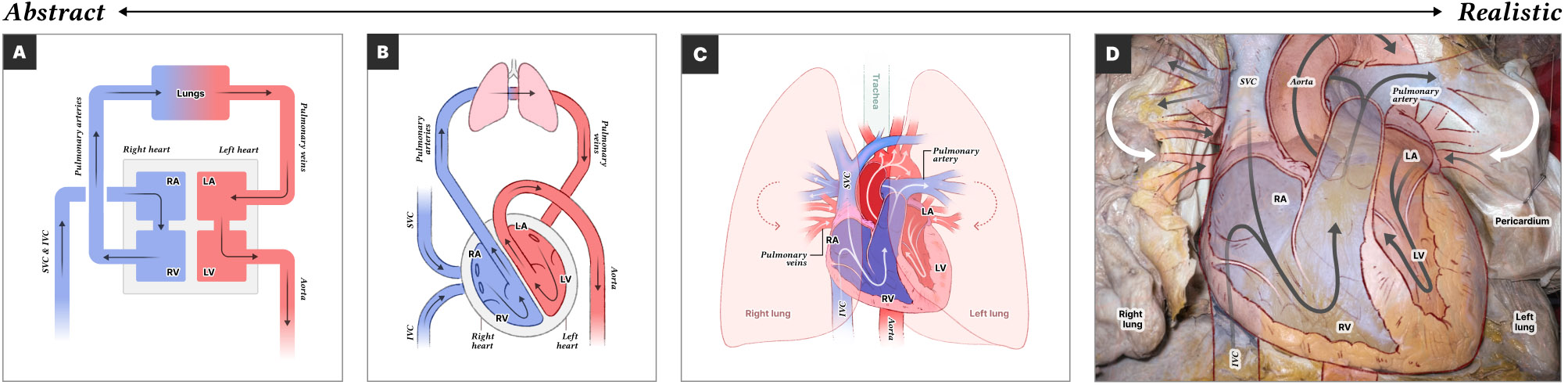}
  \caption{Didactic visualizations of the heart \& lungs that reproduce common patterns found across the literature: stimuli range from ``abstract'' (low-fidelity) to ``realistic'' (high-fidelity) but vary simultaneously across spatial, perceptual, functional, contextual, and cueing dimensions (A–D). Although visual stimuli are often described as points on a single fidelity continuum, these illustrations exemplify the idea that fidelity operates across numerous axes, complicating comparisons and interpretations. Panel D is adapted, edited, and illustrated over an original image (``Heart dissection,'' 28 October 2011), licensed under 
\href{https://creativecommons.org/licenses/by-sa/3.0/}{{\ccLogo\ \ccAttribution\ \ccShareAlike\ CC BY-SA 3.0}}, via 
\href{https://commons.wikimedia.org/wiki/File:Heart_dissection.jpg}{Wikimedia Commons}.
Panels A--C are original illustrations produced by the first author.}
  \label{fig:AbstractReal}
\end{figure*}

The community that has most directly studied how visual media affect learning is educational psychology (Section~\ref{sec:fidelityresearch}); studies frequently employ terms such as ``\textit{realistic}'' \cite{Arnold_1975, Dwyer_1968, Schwartz_1998}, ``\textit{concrete}'' \cite{Moreno, Jaakkola_2014, Zheng_2022}, ``\textit{schematic}'' \cite{Scheiter, HegartyKozhevnikov}, ``\textit{abstract}'' \cite{Cheng_2002, Jaakkola_2014, Kaminski_2008},  or ``\textit{high/low-fidelity}'' \cite{Dwyer1976, Zacharia_2011, Smallman_2010}, yet these labels rarely specify which visual features are actually being altered. The result is a persistent conflation problem: multiple representational features shift simultaneously, making it difficult to isolate causal effects or to compare and synthesize observed outcomes across studies \cite{Hffler_2007, Dwyer1976, Skulmowski_2021, Smallman_2010}. \textbf{Figure~\ref{fig:AbstractReal} reproduces a common pattern found across the literature.} The problem goes beyond comparing apples and oranges: when many dimensions shift together under the label ``\textit{more realistic},'' there is no principled basis for attributing observed effects to specific visual changes. Stimulus design for research must therefore attend carefully to these distinctions to avoid misattributing learning effects to the wrong visual feature or groupings of visual variables.

%This is not a problem confined to the learning sciences literature. Work on visual abstraction [Viola & Isenberg 2018; Viola, Chen & Isenberg 2020] establishes that abstraction operates across multiple independent axes rather than a single continuum; parallel discussions have surfaced in serious games [Rogers et al. 2022], immersive virtual environments [Huang & Klippel 2020], and 3D character visualization [Zibrek et al. 2018], where the effects of realism on users consistently depend on which specific dimensions are varied and which are held constant. Empirical work in biomedical visualization makes this concrete: Garrison et al. [2021] show that audience preferences are shaped by clarity, communicative intent, and representational convention simultaneously — factors that a single realism axis cannot disentangle. [IN BETWEEN]. This is the gap the present work addresses, specifically in the context of didactic visualization for science and engineering.

% In visualization design, even seemingly minor modifications (such as altering color intensity in a visual display) can significantly impact a viewer’s perception and subsequent mental model \cite{LOWE201672, Munzner, Johnson_Glauch, Ware}. 

% Importantly, viewers do not passively receive information; they actively construct meaning based on their prior knowledge, perceptual heuristics, and cognitive biases \cite{Schloss, Hegarty_2011, Hegarty_2014, Vekiri_2002}. 

\textbf{Concerns about these issues are not new. }Nearly five decades ago, Dwyer, one of the pioneering scholars to examine the instructional role of visualization in learning, including questions of representational realism,  articulated a challenge that remains strikingly relevant today, and shares sentiments that crosscut and motivate the current work \cite{Dwyer1976}:

\begin{fancyquote}
\textit{“Little concern is given during the production stage to the type of instructional influence the visuals will exert on the learner. }[...]\textit{ To date, very little research effort has been devoted to the \textbf{isolation, identification, classification and measurement of those essential stimuli characteristics}, both singly and in various combinations in visual illustrations }[...]\textit{”} 
\end{fancyquote}

Despite decades of ongoing research \cite{Scheiter,Goldstone_2005,Cheng_2002, Skulmowski_2023, Johnson_2014} and the unprecedented visual flexibility and sophistication afforded by contemporary tools, the core issue Dwyer identified endures. We still lack systematic ways to describe, characterize, and intentionally manipulate the representational properties of didactic visuals in science and engineering (particularly those concerning fidelity).

This gap is compounded by a structural mismatch between how practitioners and researchers conceptualize and operationalize representational fidelity. Experienced designers are highly attuned to how visual meaning emerges from countless small, interacting design decisions \cite{Ware, Munzner, Alberto} - \textit{a rendering choice here, a contextual cue there, a simplification of geometry that reshapes how a viewer reads spatial relationships}. Educational psychologists, in contrast, tend to operationalize fidelity along a continuum or into manageable categories (see Section~\ref{sec:fidelityresearch}) to make stimulus production and experimentation feasible. The result is a body of scholarly literature in which the relationship between specific visual features and learning outcomes is systematically difficult to isolate, and therefore difficult to apply. As a team of five researchers and practitioners (with backgrounds spanning visualization, design, medicine, biology, engineering, and the learning sciences), we have each encountered versions of this problem in practice: in designing didactic visualizations, in evaluation, and in teaching others to do both. 

% Our shared experiences and discussions motivating this work point to a structural disconnect in how practitioners and educational researchers conceptualize and operationalize representational fidelity. 

%  Addressing these challenges requires moving beyond monolithic notions of fidelity toward a more structured, multidimensional account.

%This absence is reflected in the broader conception of representational fidelity in scholarly work, where it remains under-specified and unevenly applied. 
Relevant theoretical foundations exist across several adjacent communities, and offer essential building blocks. Visualization research, for instance, offers theoretical accounts for reasoning about representational choices and understanding. Viola, Isenberg and Chen \cite{Ivan2018, Ivan2020} formalize visual abstraction as operating along independent axes that together span an \textit{abstraction space}; Windhager et al. \cite{windhager2024} frame complexity as a design material distributed across the visualization pipeline; Garrison et al. \cite{Garrison} found that both expert and non-expert audiences consistently preferred middle-range visual and model abstractions for biomedical process communications. Parallel concerns about the relationship between visual fidelity and comprehension appear across serious gaming, virtual environments, and other HCI contexts \cite{Risley2025, Cao2023, VisualDistraction, Rogers_2022}. These accounts offer powerful frameworks, but they do not address the specific representational demands of didactic visualization that broadly encompass science and engineering phenomena (e.g., multi-scale structure, imperceptible referents, model-mediated phenomena).

Our primary contribution is therefore a \textbf{multidimensional design space of representational fidelity in didactic visualizations for science and engineering}. In this work, we define \textbf{representational fidelity} as a multidimensional property of didactic visualizations that mediates a viewer’s access to a referent (i.e., the entity, process, or system being depicted) through design decisions that shape what is shown, how it behaves, how it is situated, what is emphasized, and how it can be explored. Fidelity here does not imply and is not limited to literal resemblance, nor does it reduce to accuracy (see Section~\ref{sec:discussion}); instead, it captures how representational choices structure interpretation and understanding across multiple dimensions. We also argue that the value of this design space lies not only in its descriptive utility, but in cultivating \textbf{Reflective Representational Reasoning} (see Section~\ref{sec:applieddesignspacesection}). Importantly, \textbf{our aim here is not to adjudicate which configurations are “best” for learning or instruction}. Instead, we treat the design space as a first step toward more principled design and scholarship. A secondary contribution lies in the corpus itself: we provide and make available the entire coded library of examples captured in this study (see \hyperref[sec:Supplemental]{Supplemental Materials}, S1).

Our analysis identifies five primary design dimensions and seven sub-dimensions (see Section~\ref{sec:designspace}, Figure 1). These dimensions were derived from a systematic examination of 175 didactic visualizations across a wide range of disciplines (including medical sciences, biological and natural systems, cellular and molecular systems, and engineering/physics) and modalities.

%This design space serves two tightly connected but often siloed audiences:\textbf{ educational researchers}, for whom it clarifies how representational fidelity can be described, compared, and operationalized in studies of visual learning. \textbf{For designers, }it offers a principled structure for reasoning about and communicating design choices related to representational fidelity. A shared conceptual structure creates common ground for more productive dialogue between these groups (see Section~\ref{sec:ApplyingDesignSpace}).

\section{Background}

\subsection{Didactic visualizations}
\label{sec:didacticvisuals}
Didactic visualizations are external visual representations designed to make a target concept, process, or system learnable through visual means \cite{Gilbert_2008, Christiansen_2022, Evagorou_2015, Jenkinson_2018, Zhang_1997}. These representations are ubiquitous across instructional contexts, including textbooks, lectures, online learning platforms, museums, etc., particularly in domains where phenomena are complex, abstract, dynamic, or otherwise inaccessible to direct experience. Visualizations can take many forms, ranging from static diagrams and annotated illustrations to animations, simulations, and interactive models \cite{Zhang2023}. In crafting didactic visuals, designers make deliberate representational choices about what to include, omit, emphasize, or simplify in service of pedagogical goals \cite{Jenkinson2017, Gilbert_2008}. These negotiations often involve trade-offs between complexity, completeness, level of abstraction, expressiveness, etc.

\textit{\textbf{A note on terminology -}} Terms such as \textit{visualization, visual display, graphical display,} and \textit{external representation} overlap in meaning across disciplines. We use them interchangeably to refer to visually mediated representations that communicate or support understanding of a target phenomenon, noting distinctions only where they matter (e.g., static vs. animated graphical display).

\subsection{Fidelity in didactic visualization}
\label{sec:fidelityresearch}
Representational fidelity is frequently treated as a defining characteristic of didactic visualizations, yet its meaning remains conceptually vague, imprecise, and inconsistent throughout scholarly works. As a consequence, simply asserting that a visualization should be higher or lower fidelity offers limited analytical, evaluative, or practical value. As part of this investigation, we began with a scan of the literature to better survey how representational fidelity is defined and applied (see \hyperref[sec:Supplemental]{Supplemental Materials}, S5). We started with sources known to the author team, followed by backward and forward citation searching across educational psychology, visualization, and HCI, resulting in a charted collection of over 30 papers; we then identified recurring definitions, dimensions, and methodological approaches relevant to representational fidelity. The results of this preliminary exercise confirmed our initial expectations; representational fidelity is characterized by substantial variability in terminology, definitions, and implementation across the literature. Many papers treat fidelity (or related conceptions like realism, abstraction, concreteness, etc.) as being self-evident, yet closer inspection reveals it is applied through different visual features, scales, and conceptual backdrops. For example, a study characterizing a visualization as \textit{higher fidelity} rarely specifies whether this rests on rendering technique, geometric detail, motion, etc. Charting the literature allowed us to trace dominant trends and notable exceptions. The vast majority of reviewed work conceptualizes fidelity as a single range, contrasting more and less realistic representations; a smaller subset problematizes this framing in more differentiated terms, an approach we extend and formalize through the design space proposed in this work. We have provided a synthesis of our observations from the literature below.

\textit{\textbf{Representational Fidelity as a range}}\textbf{ – }The prevailing framing across the reviewed literature conceptualizes representational fidelity as a single continuum, typically anchored by oppositional descriptors: high- to low-realistic detail \cite{Arnold_1975, Dwyer_1968, Schwartz_1998, Holliday_1973, Ferguson, Ware}, realistic to abstract \cite{Rieber, Joseph_1984, Gonzalez_1996, H_ffler_2010, _ltekin_2016, Ferguson}, concrete to abstract \cite{Moreno, Jaakkola_2014, Kaminski_2008, Zheng_2022}, concrete to idealized \cite{Goldstone_2005,Kokkonen_2022, Goldstone_2003}, high- to low-fidelity \cite{Dwyer1976, Zacharia_2011, Smallman_2010}, high- to low-abstraction \cite{Cheng_2002}, representative to abstract \cite{Christiansen_2022}, realistic to schematic \cite{Scheiter}, pictorial to schematic \cite{HegartyKozhevnikov}, detailed to simplified \cite{Butcher}, contextualized to abstract \cite{Johnson_2014}, high- to low-iconicity \cite{Carpenter_1953}, high- to low-perceptual information \cite{Schwartz_1995}, and perceptual richness to blandness \cite{Menendez_2020}. \textbf{These formulations converge on a shared assumption: representational fidelity can be meaningfully ordered along a single axis from \textit{more} to \textit{less}. }In instructional and learning sciences research, such continua function as convenient analytic shorthand, enabling researchers to draw tractable contrasts between sets of visual stimuli. A common rendition of this framing positions fidelity along a spectrum from \textit{more realistic} to \textit{more abstract} representations. Several frequently cited works adopt this approach, describing fidelity in terms of how closely visual depictions mirror or maintain a physical resemblance to real-world objects or phenomena \cite{Dwyer1976, Rieber, Scheiter}. Within this tradition, Christiansen situates didactic scientific visuals along a continuum that spans from figurative representations (i.e., specimen drawings) to highly abstract representations (i.e., data visualization) \cite{Christiansen_2022}. Scheiter et al. similarly describe realism as a continuum from schematic to realistic rendering, arguing that fidelity increases as more surface features of the referent are introduced \cite{Scheiter}. Hegarty further links realism to representational isomorphism, emphasizing that dynamic referents often require dynamic displays (e.g., animation), whereas static structure may be conveyed more abstractly with equal, or superior, effect \cite{Hegarty_2014}. Goldstone and Son adopt different terminology but illustrate a comparable distinction, contrasting concrete (i.e., perceptually rich) representations with more idealized, abstracted ones \cite{Goldstone_2005}. Moreno et al. similarly contrast concrete diagrams of electrical components with abstract diagrams that rely on conventional schematic symbols \cite{Moreno}.

\textit{\textbf{Representational Fidelity as multidimensional}}\textit{:} Of the reviewed scholarly works, far fewer papers define representational fidelity as a multidimensional range. For example, Höst et al. \cite{Host_2022} demonstrate that shading, color variation, and surface texture each contribute to whether images are perceived as depicting real objects. Skulmowski et al. \cite{Skulmowski_2023, Skulmowski_2021} articulate realism differently, as the combination of geometry, shading, and rendering (i.e., the GSR model). While useful for decomposing visual surface properties, the GSR model addresses only one part of what we term Morphological Fidelity (Section \ref{sec:Morpho}). Multidimensional framing also appears in other computer science, interactive systems, and HCI research where realism is often approached as an emergent system property. A systematic review of realism in digital games \cite{Rogers_2022} posits that visual realism emerges from multiple, coexisting dimensions spanning narrative, perceptual/representational and player-response factors. Slater et al. \cite{Slater} and Huang and Klippel \cite{Huang_2020} provide a description of representational fidelity within immersive virtual environments, where fidelity is described as both geometric and illumination realism. Geometric and lighting realism can be further decomposed into static and dynamic aspects, suggesting temporal coherence and behavioral consistency in sustaining realism over time. From a computer graphics perspective, these distinctions allow systems to treat geometry, lighting, real-time update mechanisms, etc. as separable components, rather than subsuming them under a monolithic notion of visual quality.

Although multidimensional accounts of fidelity exist, they remain comparatively sparse, particularly outside of computer science-driven contexts. Notably, we found no existing conception of representational fidelity that we consider sufficiently complete and well-suited to describing the design space of didactic visualizations. We build on these accounts, using the vocabulary and dimensions surfaced across disciplines as a starting point for the design space presented here.

\subsection{Design Spaces}
The concept of a design space can be understood as the landscape of conceivable design solutions, not just the ones that are obvious or currently achievable, but other configurations that could also exist. Biskjaer et al. \cite{Biskjaer_2014} define a design space as, “\textit{a conceptual space, which encompasses the creativity constraints that govern what the outcome of the design process might (and might not) be.}” Design spaces are widely employed across fields such as HCI and data visualization \cite{Solen2024,DashboardDesignPatterns,Anthropographics, Timelines,Wang_2019} to systematically explore, compare, and reason about design alternatives. In contrast, the use of design spaces in learning-focused research, particularly in the design of didactic visual materials, is often implicit. Educational design work frequently relies on convention, intuition, and craft-based knowledge \cite{Johnson_Glauch, Jenkinson2017, Zhang2023}, leaving the link between design decisions and learning outcomes difficult to articulate or examine.

\section{Methodology}
\label{sec:methods}
\begin{figure*}[t]
  \centering
  \includegraphics[width=\textwidth, alt={A flow diagram showing two methodological strands. Strand 1 depicts three iterative phases—Initialization, Expansion, and Refinement—during which the corpus grows from 38 to 175 items and the design space is developed through independent coding and collaborative review. Strand 2 shows a pilot focus group feeding into an applied design case study.}]{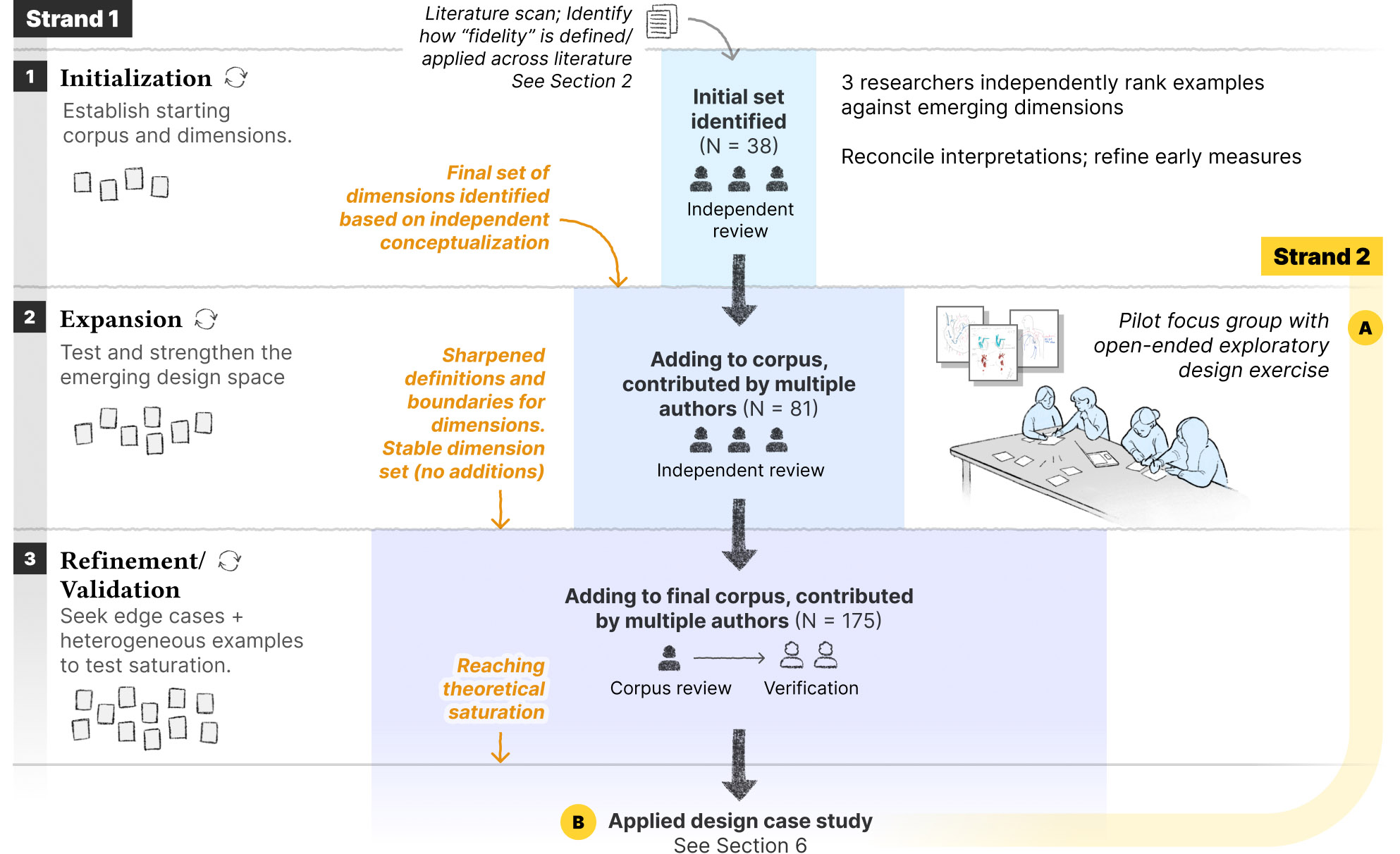}
  \caption{Study methodology overview.\textbf{ Strand 1} —The design space evolved through three phases (Initialization, Expansion, Refinement), with increasing corpus size and clarity via coding and collaborative review. \textbf{Strand 2} - The figure also highlights a pilot focus group conducted during the Expansion phase, which ultimately informed an applied design case study outlined in Section 6 as a final step in our methodology.}

  \label{fig:Method}
\end{figure*}

Section 2 identifies a clear gap: no existing account of representational fidelity adequately addresses didactic visualization. We address this gap through an iterative, multi-strand methodology \cite{Solen2024} to develop and validate the proposed design space; see Figure~\ref{fig:Method} for our methodological workflow. Strand 1 establishes the empirical foundation through a staged corpus analysis. Strand 2 complements this with a pilot focus group and applied design case study of the proposed dimensions. Together, these strands support both conceptual rigor and early practical validation of the design space. First, we outline the boundaries of the design space before describing how the corpus itself was assembled.

\subsection{Boundaries of our Design Space}
Our design space concerns \textbf{didactic visualizations }(Section~\ref{sec:didacticvisuals}). For the purposes of our corpus, this includes objects, structures, processes, or systems with a real-world referent, whether they are directly perceptible (e.g., anatomical structures), perceptible through instrumentation (e.g., engineering strain fields), or fundamentally beyond unaided human vision (e.g., molecular assemblies). We focus on science and engineering because these domains have long relied on visual representation to communicate complex, often imperceptible phenomena (Section 1), making them a rigorous testbed for this design space. This focus also reflects the motivations, expertise and research backgrounds of the author team. We include visuals for conceptual explanation, problem solving, procedural understanding, and edutainment, drawn from real-world practice, textbooks, educational platforms, and relevant scholarly research. We intentionally exclude (1) tactile, physical, or VR/AR modalities that require separate interaction design frameworks (e.g., haptic models, immersive simulations), and (2) data visualizations or statistical graphics that lack a real-world referent (e.g., abstract network graphs), as design spaces already exist for these areas \cite{Solen2024, Anthropographics, Lee_2024, AR_Benjamin, DataPhysicalization}. 

\subsection{Strand 1 - Corpus Collection and Development of Design Space}
\label{sec:strand1}

The empirical foundation for constructing the design space (as in other visualization and HCI studies, e.g. \cite{Solen2024,DataPhysicalization, Writing}) is a corpus of representational examples, assembled through a staged process. We approach our design space investigation as co-constructed by our interdisciplinary research team (5 authors; 3 served as independent coders across all phases), through literature charting, corpus building, and iterative discussions and stress-testing.

\textbf{1. Initialization}: We began with a scan of the literature as a starting point to better understand how visual fidelity is defined and operationalized in the literature (see Section \ref{sec:fidelityresearch} and \hyperref[sec:Supplemental]{Supplemental, S5}). This literature search allowed us to build a starting point for a shared vocabulary to describe representational fidelity and similar concepts, while also surfacing visual stimuli that became part of the initial corpus. In addition to these visuals, we populated our corpus with either works already known to us or otherwise emblematic cases that exemplify different divergent cases of fidelity. These included professionally oriented works and visuals used in academic investigations; a majority came from the \href{https://multimediadesignatlas.notion.site/}
{Multimedia Design Atlas} \cite{MDA}, a visual library of educational visualizations. At this stage, our aim was to cast a wide net, drawing from a diverse range of visualizations across domains, including \textbf{medical sciences, biological and natural systems, cellular and molecular systems, and engineering/physics}. Inclusion at this stage was based on the team's collective judgment that an example was emblematic of an emerging notion of fidelity, rather than a fixed set of criteria. Several authors contributed to the initial corpus.

We paused our collection at \textbf{38 works} to construct the initial dimensions, anchoring the inquiry and giving the author team a common frame for early, collaborative discussions and exploration. Through these early conversations, \textbf{six candidate dimensions were proposed; }these measures were not treated as fixed categories but as provisional constructs that would evolve through discussion and reflection. A full record of how the dimensions were introduced, revised, merged, discarded, and refined is available in \hyperref[sec:Supplemental]{Supplemental, S2}.

Following these discussions, three authors (consistent across all three corpus phases) independently explored the initial corpus, mapping the examples to the proposed dimensions and their emerging sub-dimensions. This phase was also open and exploratory. Independent coding was particularly important, allowing each contributor’s perspective (shaped by differing disciplinary experiences and media production practices) to inform the interpretation and evolution of the dimensions. In evaluating each item, researchers examined every visualization in its entirety (i.e., treating a static representation as the full set of visual elements present on the page). Textual elements embedded within the core visual were considered, while external texts (i.e., captions) were excluded (i.e., considered outside the main visual). Some multi-panel figures combined photographic and schematic elements; because photographs are not always designed with the same representational intent as diagrams, their presence was not taken to imply Perceptual Fidelity for the entire figure. To support consistency across coders, each visualization was coded for the presence or absence of each fidelity dimension and sub-dimension (i.e., binary ranking), based on whether that representational property appeared at any point within the visualization. Several rounds of discussion and reconciliation of differing interpretations followed. Through these iterations, we were able to define the dimensions more precisely, add new ones, remove others, and ultimately reach consensus across all 38 visualizations. This first phase ultimately led to the most significant refinements to the structure and definitions of the emerging design space.

\textbf{2. Expansion}: The corpus was then expanded through contributions by four authors. We were careful to include examples that could both test the strength of our initial dimensions and broaden the range of modalities, sources, and disciplines represented. At this stage, we had a total of \textbf{81} \textbf{items} in the corpus. Our analytical approach was similar to the first stage; three authors independently coded the additional visualizations with the design dimensions identified in Stage 1, with several rounds of discussion following. Any discrepancies were addressed during team meetings, with most disagreements among the three authors arising from differing interpretations and edge cases (especially around Perceptual Fidelity and Contextual Fidelity). We also revisited cases that had produced uncertainty or disagreement in the initial corpus to ensure consistency and to incorporate clarifications that emerged from subsequent discussions. The definitions were sharpened over several rounds; no new dimensions or sub-dimensions were added. 

\textbf{3. Refinement/Validation}: In the final stage, we deliberately sought out under-represented media formats (e.g., simulation-based media), and edge cases to counterbalance the biases of the expanded corpus and to test the robustness of our emerging categories. This included examples that deliberately “\textit{break}” more standard or conventional representational practices. The goal of this round was twofold: (1) to ensure heterogeneous coverage of representational practices, and (2) to reach theoretical saturation (i.e., no new dimensions needed to classify new corpus items under our operational definitions) to validate our design space. After incorporating these additional materials, the corpus reached a total of \textbf{175 items}. The first author coded all new examples, while two additional authors independently ranked a subset of 20 entries each, finding no discrepancies with the established dimensions. Importantly, no new distinctions or fidelity types emerged during this stage, and all additional examples were cleanly accommodated within the design space. This convergence strongly suggests that the design space had reached theoretical saturation. As a final verification step, the first author conducted a comprehensive audit of the entire corpus, reviewing every coded entry against the finalized dimension definitions to ensure consistency of application across all three phases of collection.

\textbf{Corpus overview}: The final corpus consists of \textbf{175 didactic visualizations }and reflects substantial variation in representational modality, source, and subject matter, providing a robust foundation for deriving and validating the design space. In terms of modality, the corpus includes 116 static visualizations (66\%), 29 interactives (17\%) and 30 animated representations (17\%). The visualizations are drawn from a range of sources, with 93 originating from professional or applied contexts (53\%), 57 from textbooks (33\%), and 25 from academic publications (14\%). Subject matter spans medical sciences (49; 28\%), engineering/physics (47; 27\%), cellular and molecular systems (44; 25\%), and biological and natural systems (35; 20\%). Collectively, this distribution captures a broad range of representational practices and reflects the types of visuals students are likely to encounter in science and engineering courses. The final corpus, including individually coded items, is made available under \hyperref[sec:Supplemental]{Supplemental Materials}, S1. A full breakdown of author contributions is provided using the CRediT taxonomy in Appendix ~\ref{sec:AppendixAuthor}, Figure~\ref{fig:authorcredit}.

\subsection{Strand 2 - Pilot Focus Group \& Design Case Study}
\label{sec:strand2}
To complement the corpus-driven development of the design space, we conducted a pilot focus group session with four second-year Master of Science in Biomedical Communications (MScBMC) students at the University of Toronto (UofT), whose training uniquely positions them to implement, interpret, and critique representational choices in didactic visualizations (see Appendix~\ref{sec:AppendixFocusGroup} for details regarding participants). The 2-hour focus group was designed to examine the interpretability and application of the proposed fidelity dimensions. We conducted this session during the Expansion phase of Strand 1 so that feedback from these skilled designer-participants could inform ongoing refinement of the design space. The session was audio recorded and transcribed; verbal rationales produced during the design exercise were analyzed in relation to the fidelity dimensions, with the first author leading analysis and a second author reviewing for consistency. Focus group materials are available under \hyperref[sec:Supplemental]{Supplemental Materials}, S3. The focus group was approved by the UofT REB (49543). Data collected from the focus group were analyzed in relation to the design space dimensions developed in Strand 1, and the latter part of the session was expanded into an applied design case study (Section \ref{sec:applieddesignspacesection}); a single-designer exploratory traversal offering a case for the design space's generative potential.

\textbf{A. Pilot focus group:} The focus group began with a brief orientation to the design space. Participants completed short preliminary exercises, including an individual categorization task (i.e., assigning dimensions to visualizations from the corpus), followed by a guided group discussion. These activities supported familiarization with the dimensions, providing context for the subsequent design-oriented portion of the session. For the remainder of the session, participants engaged in an open-ended exploratory design exercise\textbf{.} Working in groups of two, they applied the fidelity dimensions to a shared instructional task provided during the session: \textbf{blood flow through the heart}. Teams were asked to collaboratively design multiple representational variants of the same underlying phenomenon for a specified audience – \textbf{first-year anatomy or physiology students} – making their fidelity-related design choices explicit. The overarching task was to explore the design space by generating multiple representational variants that differed systematically along the fidelity dimensions; for example, altering Spatial Fidelity while holding Functional Fidelity. \textbf{This phase served two methodological purposes. }First, it assessed whether participants could operationalize the dimensions, i.e., design\textit{ with} the dimensions. Second, it provided a structured yet flexible setting in which participants could probe edge cases, surface implicit assumptions in their own design practice, and articulate tradeoffs and interactions between dimensions.

\textbf{B. Applied design case study: }Outputs from the pilot focus group (including sketches and verbal rationales) were analyzed to develop a more comprehensive design case that demonstrates a proof-of-concept implementation of the fidelity dimensions, and supports the framework’s validity and practical utility. This case study is outlined in Section \ref{sec:applieddesignspacesection}, where it serves as a worked illustration of how the design space can be applied in a real design scenario.

\section{The Design Space}
\label{sec:designspace}
The design space is organized around five primary dimensions: \textbf{Morphological Fidelity} (Spatial Fidelity, Perceptual Fidelity, and Model-derived Fidelity), \textbf{Dynamic Fidelity} (Functional Fidelity and Temporal Fidelity), \textbf{Cueing Fidelity,} \textbf{Contextual Fidelity},  and \textbf{Interactive Fidelity} (Observational Fidelity and Simulation Fidelity). Not all of the dimensions that follow are new to scholarly work or craft-based practice; terms such as Perceptual and Functional Fidelity echo related notions surveyed earlier (see Section \ref{sec:fidelityresearch}). Our contribution lies in differentiating dimensions that prior work frequently conflates, and in demonstrating their conceptual independence, synthesized into a single, comprehensive design space. Along each dimension lies a theoretically infinite continuum of potential configurations \cite{Schulz}. Inclusion in the design space does not require high fidelity along any dimension; visualizations may occupy very low positions along a dimension while still qualifying for inclusion. Figure~\ref{fig:teaser} presents an overview of the design space and its structure. Appendix \ref{sec:appendixDisciminating} provides supplementary guidance for each dimension and sub-dimension (e.g., discriminating criteria) and Appendix \ref{sec:appendixCorpusExamples} showcases several corpus examples to illustrate the range of variation across all dimensions and sub-dimensions. See \hyperref[sec:Supplemental]{Supplemental Materials}, S1,  for the entire coded library. A full record of how the dimensions were introduced, revised, merged, discarded, and refined is available in \hyperref[sec:Supplemental]{Supplemental Materials, S2}.

\subsection{Morphological Fidelity}
\label{sec:Morpho}

\textbf{Morphological Fidelity refers to the degree to which a visualization preserves or plausibly constructs the form, structure, and appearance of its referent, whether that form is directly observable or model-derived.} It encompasses the spatial organization of parts, the perceptual qualities of visible surfaces, and/or the constructed visual qualities of entities that cannot be directly perceived.
\subsubsection{Spatial Fidelity}
\noindent
\begin{wrapfigure}{l}{0.050\textwidth}
  \vspace{-1.2em}
  \includegraphics[width=0.06\textwidth]{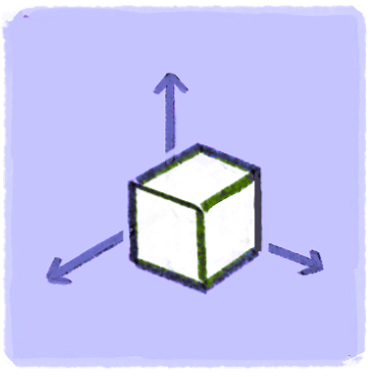}
  \vspace{-2em}
\end{wrapfigure}
\textbf{Spatial Fidelity refers to the degree to which a visualization preserves the geometric structure, relative positions, and topological relationships of its referent (see corpus entries 1, 26, and 134).} It is present when proportions, orientation, adjacency, continuity, and overall spatial arrangement correspond to the structure being represented. Importantly, Spatial Fidelity does not preclude high levels of abstraction (\textit{see corpus entry 134}), but abstraction should preserve structural information. These relationships may be maintained at varying levels of specificity, ranging from spatial cues, ordering, continuity, or connection between elements, to more literal or figurative spatial correspondences that reflect sizing, containment, and/or relative positioning. Spatial Fidelity concerns the form and how parts relate in space; surface appearance and perceptual qualities are captured by the sub-dimensions below.

\subsubsection{Perceptual Fidelity}
\noindent
\begin{wrapfigure}{l}{0.050\textwidth}
  \vspace{-1.2em}
  \includegraphics[width=0.06\textwidth]{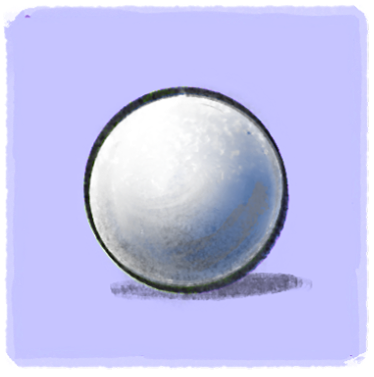}
  \vspace{-2em}
\end{wrapfigure}
\textbf{Perceptual Fidelity refers to the degree to which a visualization maintains perceptual correspondence between its visual cues and the sensory qualities of its referent (see corpus entries 16, 94, and 115).} A visualization demonstrates Perceptual Fidelity when attributes such as color, shading/lighting, material properties, etc., evoke an appearance consistent with how the referent would actually look or be perceived at the represented scale. \textbf{Perceptual Fidelity applies only when the referent possesses an observable appearance}, either directly (e.g., anatomical structures, everyday objects) or through extended human perception (e.g., microscopic or telescopic imaging, where instruments merely amplify sensory access). Perceptual Fidelity does not meaningfully apply when the referent lies outside the range of human perceptual experience, such as subcellular machinery, forces, or astronomical phenomena; these entities have no true appearance to be reproduced. In such cases, visual attributes are better understood as model-derived conventions (\ref{sec:ModelDerived}) and/or cueing strategies (\ref{sec:Cueing}), not perceptual correspondence.

\subsubsection{Model-derived Fidelity}
\noindent
\begin{wrapfigure}{l}{0.050\textwidth}
  \vspace{-1.2em}
  \includegraphics[width=0.06\textwidth]{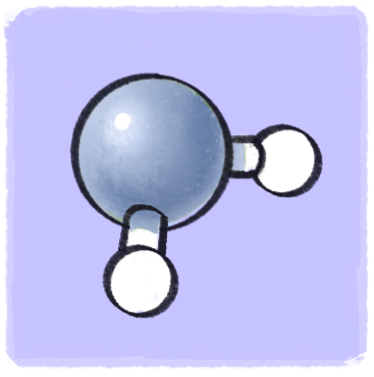}
  \vspace{-2em}
\end{wrapfigure}
\label{sec:ModelDerived}
\textbf{Model-derived Fidelity refers to the degree to which a visualization constructs a coherent and consistent sense of appearance for an entity or phenomenon that cannot be directly observed, using visual attributes that simulate perceptual qualities without claiming perceptual veracity (see corpus entries 28, 81, and 150).} Model-derived Fidelity is present when design choices (such as color, shading, lighting, material properties, etc.) are used to create a plausible, interpretable, and/or conventionally accepted appearance for entities whose true perceptual qualities are fundamentally imperceptible to humans (e.g., subcellular machinery, forces, astronomical phenomena, etc.). The constructed appearance may be informed by scientific data, computational models, disciplinary norms, or established visual conventions. Model-derived Fidelity is therefore distinct from Perceptual Fidelity. \textbf{It does not aim to reproduce how something looks; instead, it provides a \textbf{visual proxy} that supports reasoning, communication, or learning about imperceptible structures or processes.}

\subsection{Dynamic Fidelity}
\textbf{Dynamic Fidelity refers to representational choices concerning how a visualization depicts change or behavior over time.} It is organized into two sub-dimensions: Functional Fidelity and Temporal Fidelity. These two dimensions capture the representational fidelity of system dynamics, distinguishing visualizations that simply describe structure from those that communicate temporal information and/or behavior. 
\subsubsection{Functional Fidelity}
\noindent
\begin{wrapfigure}{l}{0.050\textwidth}
  \vspace{-1.2em}
  \includegraphics[width=0.06\textwidth]{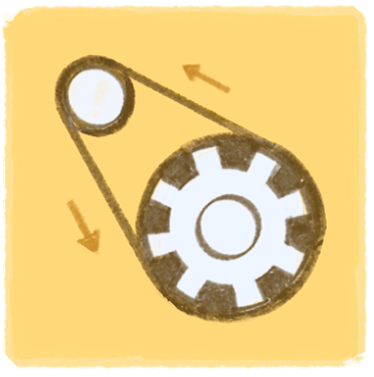}
  \vspace{-2em}
\end{wrapfigure}
\textbf{Functional Fidelity refers to the degree to which a visualization represents the causal and mechanistic relationships within a system, i.e., how one component, action, or variable produces change in another (see corpus entries 27, 77, and 131).} It is present when the visualization allows viewers to understand, infer, or predict how the system behaves when conditions change. It requires encoding how or why one state leads to the next. For example, an animation showing how changes in load redistribute stress across a structure (see corpus entry 75) demonstrates higher Functional Fidelity than one that simply juxtaposes before and after states without representing the causal linkage.

\subsubsection{Temporal Fidelity} 
\noindent
\begin{wrapfigure}{l}{0.050\textwidth}
  \vspace{-1.2em}
  \includegraphics[width=0.06\textwidth]{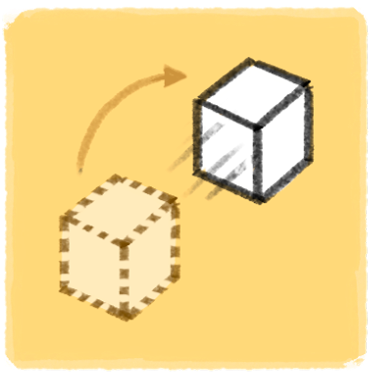}
  \vspace{-2em}
\end{wrapfigure}
\label{sec:Temporal}
\textbf{Temporal Fidelity refers to the degree to which a visualization encodes change over time, whether through animation, or through ordered sequences, staged progression, or states that encode temporal information (see corpus entries 12, 109, and 142).} At minimum, Temporal Fidelity requires that two or more temporally distinct states are depicted with an interpretable sense of before and after. Temporality may be represented as linear progression, and cyclical or periodic behavior, such as wingbeat cycles (see corpus entry 13), respiratory rhythms (see corpus entry 9), or seasonal oscillations (see corpus entry 14). It is absent when multiple states are shown with no implied ordering or progression. Importantly, viewpoint changes (e.g., camera rotations) enhance Spatial Fidelity by exposing structure from multiple angles, but do not encode temporal change unless the system itself transforms over time.

\subsection{Cueing Fidelity}
\noindent
\begin{wrapfigure}{l}{0.050\textwidth}
  \vspace{-1.2em}
  \includegraphics[width=0.06\textwidth]{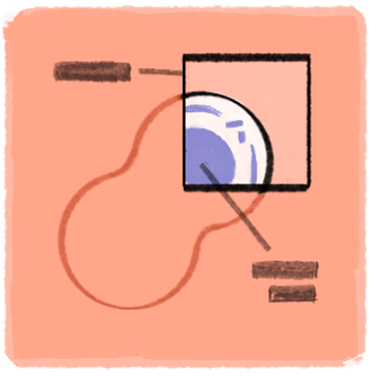}
  \vspace{-2em}
\end{wrapfigure}
\label{sec:Cueing}
\textbf{Cueing Fidelity refers to the degree to which a visualization guides the viewer’s attention and perceptual focus through visual emphasis (see corpus entries 9, 97, and 155).} A visualization demonstrates Cueing Fidelity when perceptual variables (color, contrast, motion, opacity, labels/annotations, etc.) are used to direct attention toward specific information/relationships. A common strategy involves reducing Perceptual Fidelity globally while applying strong cueing locally (i.e., a muted rendering serves as a substrate upon which specific elements are foregrounded). A notable edge case arises with disciplinary conventions: for instance, in medical illustration, veins rendered in bright blue serve as a visual code for differentiation, exaggerating a real perceptual property (cooler-hued deoxygenated blood) in service of salience. Akin to Model-derived Fidelity (Section \ref{sec:ModelDerived}), Cueing Fidelity does not require direct resemblance to a referent's appearance; it instead mediates which aspects of a referent a viewer engages with or attends to (Section 1).

\subsection{Contextual Fidelity}
\noindent
\begin{wrapfigure}{l}{0.050\textwidth}
  \vspace{-1.2em}
  \includegraphics[width=0.06\textwidth]{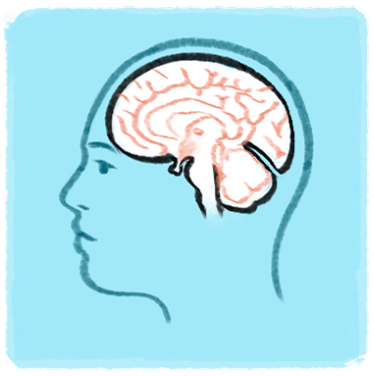}
  \vspace{-2em}
\end{wrapfigure}
\textbf{Contextual Fidelity refers to the degree to which a visualization provides contextual information that situates the represented subject within a surrounding frame, environment, or reference system (see corpus entries 5, 64, and 90).} It is present when the visualization conveys the conditions, surroundings, or relational anchors that define where, how, and in relation to what a phenomenon exists or operates. Context may include, but is not limited to, anatomical settings, geographic locations, environmental conditions, laboratory or experimental configurations, coordinate systems, spatial orientations, and/or scale references. Importantly, context does not need to be realistic or complete to be present. It may be stylized, simplified, abstracted, or partial. A special case arises when the subject itself is a contextual/reference system (e.g., anatomical planes used as a reference frame, see corpus entry 55); in such cases, a visualization exhibits Contextual Fidelity when the figure’s primary function is to establish or preserve a reference system. Context can also emerge from the relationships among elements rather than from environmental context alone (see corpus entries 11 and 25). 
%This effect is strongest when one element is clearly established as the primary subject or main actor of the scene, and the remaining elements function as comparative, supportive, or situating references (see corpus entry 11).

\subsection{Interactive Fidelity}
\textbf{Interactive Fidelity refers to the degree to which a visualization affords meaningful user engagement with the represented system, whether through exploratory actions or manipulative interventions. }It captures how well the interactive structure supports users in accessing, revealing, or altering information in ways that deepen understanding of the entity and/or phenomenon. Importantly, Interactive Fidelity is not achieved by the mere presence of interactive components.

%; interactions must expose structure, logic, relationships, mechanism, etc., for the entity or phenomenon. 

\subsubsection{Observational Fidelity}
\noindent
\begin{wrapfigure}{l}{0.050\textwidth}
  \vspace{-1.2em}
  \includegraphics[width=0.06\textwidth]{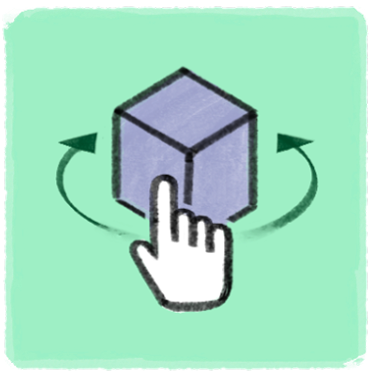}
  \vspace{-2em}
\end{wrapfigure}
\textbf{Observational Fidelity refers to the degree to which a visualization supports interactive exploration through operations that allow the viewer to observe, reveal, or inspect different aspects of the representation (see corpus entries 26, 38, and 147)}. A visualization exhibits Observational Fidelity when users can manipulate the view or presentation in ways that expose structure, spatial relationships, internal organization, or comparative detail that cannot be accessed from a single viewpoint. Typical observational interactions include rotating, zooming, panning, slicing or cutting planes, toggling layers, isolating components, highlighting elements, adjusting opacity, or switching representational modes. 
%These interactions do not alter the system itself; rather, they alter how the system is observed.
\subsubsection{Simulation Fidelity}
\noindent
\begin{wrapfigure}{l}{0.050\textwidth}
  \vspace{-1.2em}
  \includegraphics[width=0.06\textwidth]{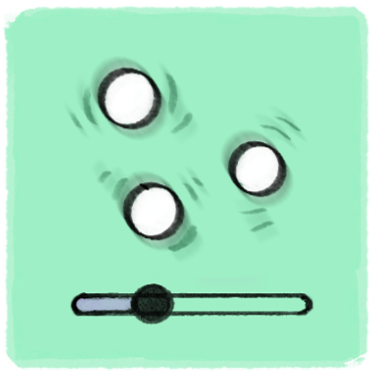}
  \vspace{-2em}
\end{wrapfigure}
\textbf{Simulation Fidelity refers to the degree to which a visualization allows the viewer to manipulate parameters, conditions, or variables in ways that produce consequential changes within the represented system (see corpus entries 27, 80, and 130)}\textbf{, whether those changes are computed through real-time simulation, or precomputed/prerendered model states}. A visualization exhibits Simulation Fidelity when user-driven changes generate outcomes that reflect the underlying causal, mechanistic, or mathematical logic of the modeled phenomenon. The defining criterion is that interactions modify the system, not merely the viewpoint. Examples include adjusting molecular concentrations to observe changes in reaction rates (see corpus entry 27) or applying forces to a mechanical system to observe resulting motion (see corpus entry 80).

\section{Analysis of Design Dimension Distributions}
\label{sec:analysis}

We analyzed the distribution and co-occurrence of the proposed dimensions across the full corpus of 175 items to characterize the structure of the design space, revealing patterns, affinities, and tensions. Although assembled through purposeful rather than systematic sampling (see Section \ref{sec:discussion}), it remains a useful lens for examining common patterns in existing visualization designs. The detailed distribution data and the correlation matrix are provided in Appendix \ref{sec:appendixCorrelation} (Figure~\ref{fig:corpusanalysis}).

Two dimensions - Spatial Fidelity and Cueing Fidelity - appear in all 175 corpus items. We interpret this ubiquity as a key finding: some degree of interpretable spatial organization and attentional guidance appears to function as a baseline requirement. The communicative and generative value of Spatial and Cueing Fidelity is demonstrated directly in our applied case study (Section \ref{sec:applieddesignspacesection}, Appendix \ref{sec:appendixAppliedDesignCase}), where both dimensions functioned as active levers for ideation rather than fixed defaults. Contextual Fidelity appears in 142 items, indicating that most visualizations situate their subject within a surrounding frame or reference system. Functional Fidelity is also common in our corpus (139 items), reflecting a strong emphasis on causal and mechanistic explanation. Perceptual Fidelity (110 items) and Temporal Fidelity (95 items) are present in just over half of the examples. Model-derived Fidelity appears in 81 items. Interactive Fidelity is comparatively less common in our corpus: 29 items employ Observational Fidelity, and 12 employ Simulation Fidelity. Although our sampling was selective and purposeful, this pattern likely reflects both the dominance of static media in instructional contexts and the higher implementation costs associated with interactive and simulation-based representations.

Pearson correlation coefficients, computed over binary presence/absence coding, quantify the degree of independence and association among dimensions
(see Figure~\ref{fig:corpusanalysis}, Panel B for details). Perceptual Fidelity and Model-derived Fidelity show a strong negative relationship (r = –0.54), reflecting a fundamental referential distinction: when a phenomenon has an observable appearance, designers tend not to rely on constructed visual proxies, and vice versa. Importantly, these oppositions are not absolute: for instance, our corpus exhibits a number of visualizations that use multi-panel or hybrid designs that deliberately integrate perceptual information with model-derived representations (see corpus entries 29, 58, 97, 159). Functional Fidelity and Temporal Fidelity are moderately correlated (r = 0.47), suggesting that depictions of system behavior often, but not always, co-occur with explicit representations of change over time. Contextual and dynamic dimensions exhibit weaker and more selective relationships. Contextual Fidelity is positively associated with Perceptual Fidelity (r = 0.32) but negatively associated with Model-derived Fidelity (r = –0.20). Within Interactive Fidelity, Observational and Simulation Fidelity are strongly associated (r = 0.61), indicating that simulation-based interactions typically build on inspection-oriented tools. Simulation Fidelity also shows modest associations with Temporal (r = 0.25) and Functional Fidelity (r = 0.14), consistent with its reliance on system dynamics, but these relationships are far from deterministic.
These patterns suggest that the dimensions exhibit characteristic affinities and tensions, some naturally co-occurring, others standing in partial opposition.

\section{Design Space in Practice}
\label{sec:applieddesignspacesection}

\textbf{In this section, we extend the insights from the pilot focus group into an applied design case study. The goal is twofold: to demonstrate how the design space is actively traversed during real design work, and to draw out broader implications for research and practice.}

During the focus group, participants engaged in an open-ended exploratory design exercise, producing sketches and verbal rationales that revealed how designers reason about and apply representational fidelity in practice. We developed an applied design case study that operationalizes these findings. 
The first author (\textit{who is an experienced visualization practitioner}) conducted the case study using the same prompt employed in the pilot focus group (\textit{\textbf{blood flow through the heart}}), navigating the design space through extended, self-directed exploration, moving from an intuition-driven baseline through a structured ideation phase in which the fidelity dimensions were actively engaged as generative tools. As a single-designer case study, this is best read as a proof-of-concept illustration rather than independent evidence of generative power of the design space (Appendix \ref{sec:appendixAppliedDesignCase}). Domain familiarity was a deliberate requirement for both the focus group and applied case study; reasoning about representational fidelity requires sufficient knowledge of the referent to evaluate what a visualization is preserving, constructing, or eliding. \textbf{A complete, annotated account of the full traversal, including all design variants with dimensional annotations, is available in Appendix \ref{sec:appendixAppliedDesignCase} (Figure~\ref{fig:DesignCaseStudy}).}

\textit{\textbf{Reflective Representational Reasoning }}\textbf{–} A central pattern emerging from both the focus group and the applied case study is the deliberate use of the design space as a tool for exploration and ideation. Participants described using the design dimensions to intentionally generate multiple variants that follow different paths through the space, treating initially unsuccessful designs as starting points for adjustment rather than discarding them. As one participant put it, \textit{“Even though some things obviously don’t work, maybe with a little more adjustment, it could become something accidentally really, really great.”} Several framed the dimensions explicitly as an aid to ideation; one participant described the process as an invitation to \textit{“explore the design space fully…push yourself to the limits…increase fidelity here, decrease there…uncover trade-offs.”} Importantly, this did not replace existing design habits. Initial sketches were still driven by the goals of the communication challenge, with the fidelity dimensions entering later to interrogate and extend those decisions: as one participant put it, \textit{“The initial sketches still start more with the goals of the drawing…then after that, you can tweak these [design dimension] sliders to explore that space further.”} Ultimately, the dimensions seemed to become structured provocations that encouraged participants to think beyond habitual patterns, generate unconventional variants, and identify tradeoffs that may otherwise remain underexplored.

Based on these preliminary findings, the design space appears to function as a reflective aid. It supports what we describe as \textbf{Reflective Representational Reasoning}: the practice of deliberately and explicitly reflecting on how representational fidelity dimensions (and their interactions) structure a viewer's access to meaning, emphasis, and understanding. By naming dimensions that are often left implicit, the framework gives designers a structure within which to experiment, helping them reason about why certain choices feel appropriate and what might be gained or lost by pushing those choices in different directions \cite{Bardzell, Sch_n_2017, Sedlmair}. Another interesting observation from the pilot focus group was the potential utility of the framework to support novice designers: \textit{“the design space could be useful for early medical illustrators who are just learning how to ideate [...] and discover what is effective.” }Participants also highlighted the utility of the design space as a useful communication tool. As one participant put it, “\textit{These [design dimensions] are things we are all always tweaking… but the fact that we now have words to describe them is really helpful}.” Having a shared vocabulary made it easier for participants to negotiate disagreements and explain design decisions that would otherwise remain tacit. This effect was not solely on communication: several participants described using the dimensions to generate variants they had not initially considered, e.g., to "\textit{push yourself to the limits... uncover trade-offs}," suggesting generative as well as communicative value.

\begin{fancyquote}
\textbf{For practitioners}, the fidelity dimensions provide a structured framework for articulating representational decisions often made tacitly. Rather than replacing intuitive design processes, the design space can be used reflectively, to interrogate and extend initial judgments. Designers may begin with a baseline representation, then deliberately vary one or two dimensions while holding others stable. Insights from the focus group and case study suggest the design space supports two distinct functions: more effective critique and communication of representational choices, and more deliberate generation of alternatives. Preliminary evidence also suggests the framework may be valuable for novice practitioners, offering a structured entry point for reasoning about representational fidelity.
\end{fancyquote}

\begin{figure}[htbp]
  \centering
  \includegraphics[width=0.48\textwidth, alt={A two-axis chart showing how combinations of Spatial and Perceptual Fidelity define distinct regions of the design space. Several example visualizations from the applied case study are placed within the space to illustrate different combinations of these two dimensions.}]{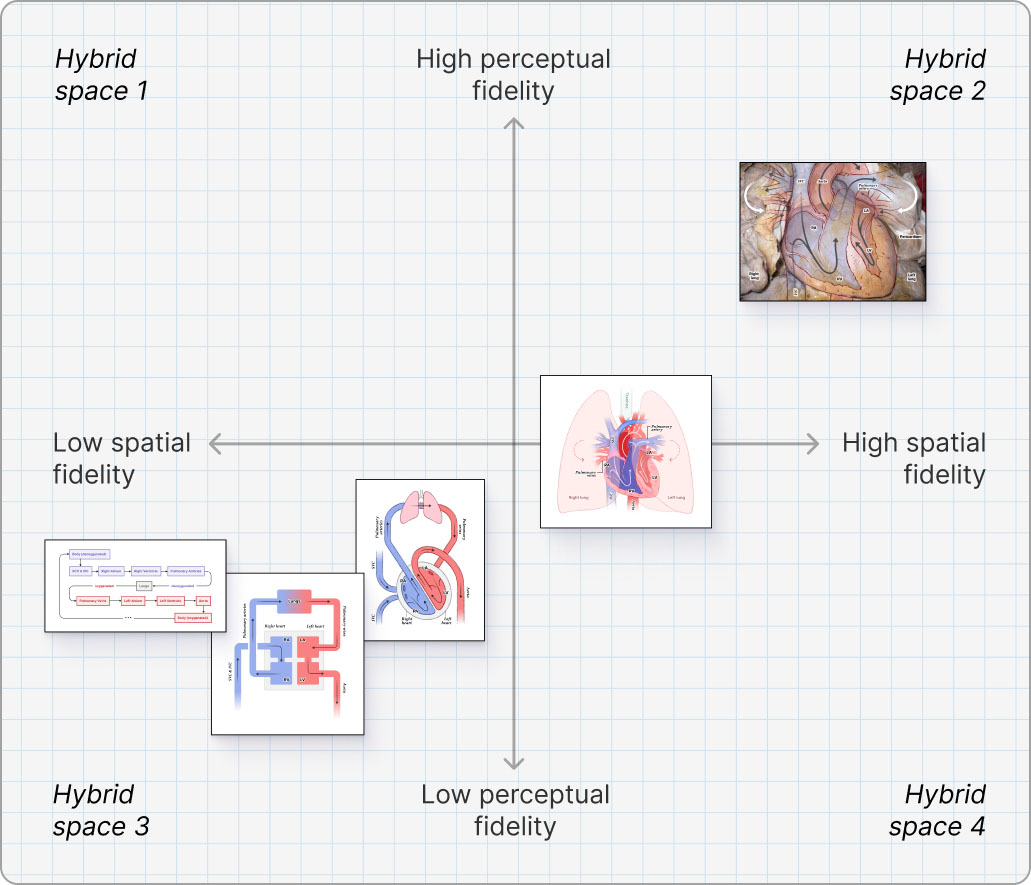}
   \caption{Relationship between Spatial Fidelity and Perceptual Fidelity, showing how their combinations define distinct regions of the representational design space. Five representative examples from the applied design case study are shown in approximate locations. Only two fidelity dimensions are visualized for clarity; other dimensions (e.g., Contextual Fidelity) are also varied in the examples but are not shown in this simplified projection.}

  \label{fig:HybridSpace}

\end{figure}

\textbf{Interaction and compounding effects of the design space –} Although individual dimensions are separable, and can be adjusted independently, combinations produce compounding effects whose impact cannot be inferred. In this sense, combining fidelity dimensions gives rise to higher-order effects in which the whole exceeds the sum of its parts, an observation that directly motivates the need for a multidimensional design space. The applied design case study illustrates this interaction; we showcase this effect by mapping the generated design variants along only two dimensions, Spatial Fidelity and Perceptual Fidelity, which produces four distinct spaces (Figure~\ref{fig:HybridSpace}). As illustrated earlier in Figure ~\ref{fig:AbstractReal} (where several dimensions differ at one time), it becomes difficult to isolate which specific representational factors (or combinations of factors) were responsible for the observed effects. Without an explicit accounting of interacting dimensions, this ambiguity can undermine internal validity and limit the interpretability of results.

\begin{fancyquote}
\textbf{For researchers,} the design space offers a means of operationalizing representational fidelity with greater specificity. Rather than relying on global labels such as \textit{realistic} or \textit{abstract}, researchers can explicitly describe which fidelity dimensions are manipulated, which are controlled, and how these choices relate to the pedagogical aims of the work. This design space can also be applied retrospectively, as an analytic lens for examining prior work. Recharacterizing existing experimental stimuli in terms of fidelity dimensions may help explain divergent and, inconsistent findings in the literature. Importantly, the design space foregrounds the possibility of interaction effects among dimensions, an issue we return to in the Discussion (Section \ref{sec:discussion}).
\end{fancyquote}

\section{Discussion}
\label{sec:discussion}
The proposed design space offers a conceptual structure for clearer reasoning, communication, and experimentation in the design and study of didactic visualizations. Following conventions in design-space research (e.g., \cite{Solen2024}), we assess our framework through its descriptive, generative, and evaluative power. We also discuss the strengths and boundaries of the proposed design space, its current and potential evaluative role and the open questions that remain. We also clarify the limitations of our corpus, our methodological choices, and the scope of claims we believe the present work can responsibly make.

\subsection{Descriptive, Generative, and Evaluative Power}

\textbf{Descriptive Power - }The design space exhibits strong descriptive power. All dimensions and sub-dimensions emerged through iterative corpus analysis and remained stable through successive rounds of expansion and refinement. Each dimension captures a distinct aspect of representational fidelity necessary for differentiating the examples in our corpus. Importantly, the design space reached theoretical saturation: after the initial set of examples had been analyzed, no new dimensions or sub-dimensions were required to account for any of the remaining items. Newly added examples sharpened definitions but did not expand the structural boundaries of the space. Likewise, the pilot focus group and applied case study demonstrate the framework’s forward-looking descriptive power: visualizations beyond the original corpus can be systematically classified, newly created representations can be described in a stable and consistent manner, and the dimensions provide a shared vocabulary for articulating and justifying design decisions.

\textbf{Generative Power - }The design space also demonstrates preliminary generative potential. Our corpus analysis offers a descriptive account of how representational dimensions are distributed and combined in existing designs, revealing characteristic patterns, affinities, and tensions. Our pilot focus group and subsequent applied case study (single-designer) further illustrate the design space’s capacity to support idea generation. In both contexts, the framework enabled designers to explore representational strategies and to articulate design choices more explicitly. We expect this utility to be especially valuable for less experienced designers and interdisciplinary teams lacking a shared vocabulary for representational decisions, where the design space can function as a scaffold for ideation, helping teams negotiate trade-offs, surface assumptions, and explore a wider range of options.

\textbf{Evaluative Power -} We do not yet claim evaluative power for this design space. We offer only preliminary, descriptive support for evaluative use, based on our systematic corpus analysis, pilot focus group, and applied design case study. We leave further efforts to develop, test, and validate more robust evaluative applications of the design space.

\subsection{Limitations}
We acknowledge several limitations that bound the scope of our claims.

\textbf{Sample-of-Convenience and Intentional Biases:} The corpus was assembled through selective and purposeful sampling, rather than systematic retrieval. The initial stages were built from sources familiar to the authors: professional examples, academic stimuli, widely circulated online materials, etc. This inevitably introduces sampling bias, but in ways that were partially intentional: our aim was to stress-test the design space using edge cases, outliers, unconventional representations, and examples spanning diverse disciplines and media forms. We do not treat the final corpus as a representation of \textit{actual} instructional materials used across scientific and engineering education. Rather, the corpus should be understood as a provocative, diverse testbed designed to probe the conceptual boundaries of representational fidelity and ensure the framework could accommodate heterogeneous practices.

\textbf{Focus Group Limitations:} The focus group consisted of second-year MScBMC students, highly trained scientific visualizers with deep domain knowledge and sophisticated representational intuitions (see Appendix \ref{sec:AppendixFocusGroup} for details). Their expertise yielded rich insight into how skilled practitioners interpret and operationalize the dimensions, but it also limits generalizability: educators, learning scientists, and novice visualization designers may interpret certain dimensions differently or find different points of friction. Increasing the sample size and engaging with a larger, more diverse group of participants will therefore be necessary to assess how broadly interpretable and usable the design space is across scientific and engineering domains, experience levels and professional contexts. 
Another related limitation concerns the\textbf{ tools and technologies }available during the focus group. The exploratory design activities relied primarily on thumbnailing (i.e., small low-detail sketches), which supports rapid ideation but constrains the expression of ideas that depend on advanced media affordances. While thumbnailing does not inherently restrict conceptual exploration, it may limit how designers ideate about, realize, and communicate certain aspects of fidelity. Time constraints for the focus group may have also influenced the depth and breadth of exploration, though the extended applied design case study (Section 6) partially compensates by demonstrating how the dimensions function under sustained, self-directed engagement.

\textbf{Cultural and Linguistic Bias: }All examples in the corpus are drawn from English-language sources, and the vast majority were produced within Western institutional and pedagogical contexts, where conventions of visualization follow particular historical, aesthetic, and epistemic lineages \cite{d2023data, Saharan}. Representational norms in other educational cultures, visual traditions, or linguistic communities may introduce fidelity distinctions not captured here, or may emphasize different relationships among dimensions. Expanding the corpus is therefore an important direction for future work.

\subsection{Open Questions \& Future Work }

While we focus here on defining the structure and practical utility of the proposed design space, several open questions remain regarding its practical uptake, pedagogical implications, and broader evaluative scope. These questions point toward a rich agenda for future research.

\textbf{Adoption in Practice}: A central open question concerns whether, and how, this design space meaningfully shifts representational decision-making in real-world contexts. In particular, it remains unclear how the design space functions within craft-based, intuition-driven workflows that dominate visualization practices \cite{Jenkinson2017}. This raises several applied questions: \textit{Can the design space be adopted through lightweight prompts, checklists, or design heuristics without imposing excessive rigidity? Does it support faster ideation, clearer justification of design choices, or the discovery of novel and interesting representational solutions? Might it also support designer–client communication, influencing how representational choices are negotiated, justified, and contested in real-world project settings?} Exploring these questions through co-design with a range of practitioners (and clients) may help surface the framework’s practical affordances and limits, as well as inform the development of tools that preserve flexibility while supporting Reflective Representational Reasoning about fidelity.

\textbf{Adoption in Research}: Likewise, an important direction lies in understanding how educational researchers (particularly educational psychologists, learning scientists) interpret and apply the design space. Future work should investigate whether the framework supports reasoning about stimulus design and selection, and whether it influences how researchers (re)conceptualize representational fidelity in relation to visual stimuli design. A related methodological question concerns granularity: both our focus group and case study surfaced reasoning about small incremental adjustments along individual dimensions, suggesting that each dimension could ultimately be operationalized along a graduated scale (e.g., Likert-type ratings) to enable more precise stimulus characterization and cross-study comparison. In addition to prospective use, an open question is whether the design space can function productively as a retrospective analytic lens, recharacterizing stimuli from prior studies to help explain divergent or conflicting results in the literature, including potential interaction effects among fidelity dimensions. This work may also help bridge the gap between research and practice by making design decisions more transparent and actionable for educators.

\textbf{The Pedagogical Layer:} In constructing the design space, we deliberately avoided embedding pedagogical intent directly into its dimensions. This decision reflects a commitment to describing representational structure rather than prescribing instructional goals. This choice leaves open compelling, albeit herculean, opportunities to map pedagogical intent alongside representational fidelity. Such an extension invites exploration of how fidelity interacts with learners’ prior knowledge, instructional context, learning goals, etc. \cite{Kalyuga_2007, MayerRichard, Hegarty_2014}.

\textbf{Accuracy and Representational Fidelity:} A closely related and equally compelling question concerns the relationship between fidelity and accuracy. While accuracy is often treated as a prerequisite for effective instructional visuals, its interaction with representational fidelity remains under-theorized. Future work should clarify how accuracy operates within the design space, and whether different forms of fidelity place distinct demands on what counts as accurate within different contexts. For example, when instructional goals prioritize conceptual understanding over surface detail, higher accuracy along some dimensions may be less productive than carefully chosen departures from it. Clarifying how accuracy and fidelity interact may help explain why representations that are technically correct do not always support learning, and why other representational configurations/strategies sometimes do.

\textbf{Interaction with Affect, Ethics, and Trust}: The design space also invites examination of fidelity through affective and ethical lenses, as well as its interaction with learners’ sense of trust and safety in instructional materials. For example, heightened Perceptual Fidelity in a surgical visualization may support understanding for trained clinicians, yet simultaneously provoke anxiety, discomfort, or disengagement among novice learners, patients, or the general public \cite{mader2005medical, Haragi_2019, Coulter_2024}. Future research should therefore evaluate fidelity from pluralistic viewpoints; we argue that the proposed design space provides a useful conceptual backdrop for such investigations.

\textbf{Context-Specific Extensions of the Design Space:} While this work targets didactic visualizations broadly, future studies may benefit from scoping the design space to more specific contexts, such as discipline-specific textbooks, assessment materials, or professional training media. Narrowing the domain would allow for systematic sampling strategies, inter-rater reliability analyses, and quantitative examinations of representational patterns.

\textbf{Within-Dimension Variation Across a Single Stimulus:} Our binary coding (Section \ref{sec:strand1}) does not capture cases where a single dimension varies in degree across a stimulus, e.g., a low-fidelity contextual element beside a larger, high-fidelity focal element within the same image. Characterizing this kind of within-dimension variation, particularly for evaluative and experimental use where precise stimulus control matters, remains an open question we leave for future work.

\section{Conclusion}
In this paper, we introduce a multidimensional design space of representational fidelity for didactic visualization, comprising five dimensions and seven sub-dimensions, derived through iterative analysis of a 175-item corpus and informed by a pilot focus group and applied design case study. The resulting framework demonstrates strong descriptive power and generative potential, facilitating clearer articulation of representational choices and revealing design patterns and opportunities present across contemporary didactic media. Our efforts lay the groundwork for future research, including systematic experiments that manipulate fidelity dimensions in controlled ways, investigations into pedagogical alignment, and utility in design practice. This framework is not a final statement but a foundation, one we invite the community to test, critique, refine, and extend.

\section*{Supplemental Materials}
\label{sec:Supplemental}

The supplemental materials accompanying this paper are made available on the Open Science Framework (OSF) to
support transparency, reproducibility, and reuse: \href{https://osf.io/rx749/}{Link to Supplemental Materials on OSF}.

%% if specified like this the section will be omitted in review mode

\bibliographystyle{abbrv-doi-hyperref}

\bibliography{02-sample-base}

\clearpage

\appendix % You can use the `hideappendix` class option to skip everything after \appendix
\crefalias{section}{appendix} % this is to make sure that cleverref switches to referring to Appx. X from here on

% \newpage

\section{Author Contribution}
\label{sec:AppendixAuthor}
\begin{figure}[htbp]
  \centering
  \includegraphics[width=0.47\textwidth, alt={A matrix table displays contributor roles and author contributions. Rows correspond to CRediT roles and columns correspond to individual contributors. Solid black cells indicate primary or substantial contributions to a role, while solid gray circles indicate secondary contributions. Roles not applicable to this work, including Software and Funding acquisition, are not shown.}]{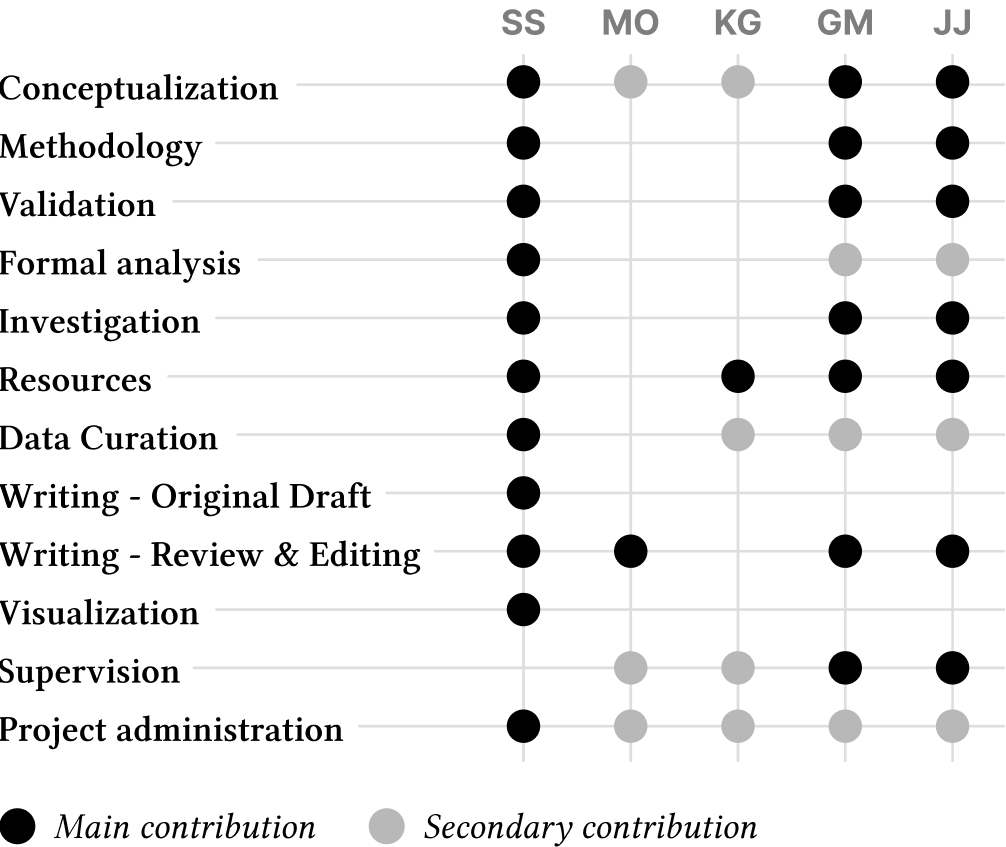}
  \caption{Authorship contribution matrix using the CRediT taxonomy. Columns represent contributors and rows represent roles. Solid black cells indicate main/substantial contributions to the corresponding role, and solid gray circles indicate secondary contribution. CRediT roles not applicable to this work (Software, Funding acquisition) have been omitted.}
  \label{fig:authorcredit}
\end{figure}

Author contributions are summarized using CRediT (Contributor Roles Taxonomy). A visual authorship contribution matrix is provided in (Figure~\ref{fig:authorcredit}) to improve transparency of contributor roles.

\section{Focus Group Participants}
\label{sec:AppendixFocusGroup}
\textbf{Participant Background}: The focus group consisted of four second-year MScBMC students at the University of Toronto (UofT), enrolled in a graduate program focusing on medical illustration. This program combines intensive scientific training with graduate-level instruction in the visual communication of science and the design and production of didactic visualizations across multiple modalities. By their second year, students have completed much of their scientific-focused training alongside studio-based coursework in illustration, animation, and interactive media. The study required participants who could engage critically with the fidelity dimensions, not only as designers, but as scientifically informed ones: able to identify when a representational choice departs from biological or physical reality, articulate why that departure might be pedagogically motivated, and surface ambiguities in the dimensions that a less domain-literate participant might not notice.

\textbf{Recruitment}: Participants were recruited from a cohort not directly supervised by any member of the research team, to minimize the potential for social desirability effects or perceived pressure to respond favorably. Recruitment was conducted via an open call distributed within the program. Participation was voluntary, and participants were informed that their feedback would be used to refine a research framework under development. No incentives were offered. Participants had no exposure to the proposed design space before the session. The focus group was reviewed and approved by the UofT REB (No. 49543). All participants provided written informed consent prior to the session.

\begin{figure}[!ht]
  \centering
  \includegraphics[width=0.47\textwidth, alt={A diagram illustrating the progression from a pilot focus group with an open-ended design exercise, through analysis of participant sketches and verbal rationales, to the development of an applied design case study. Sample participant sketches are shown below the flow.}]{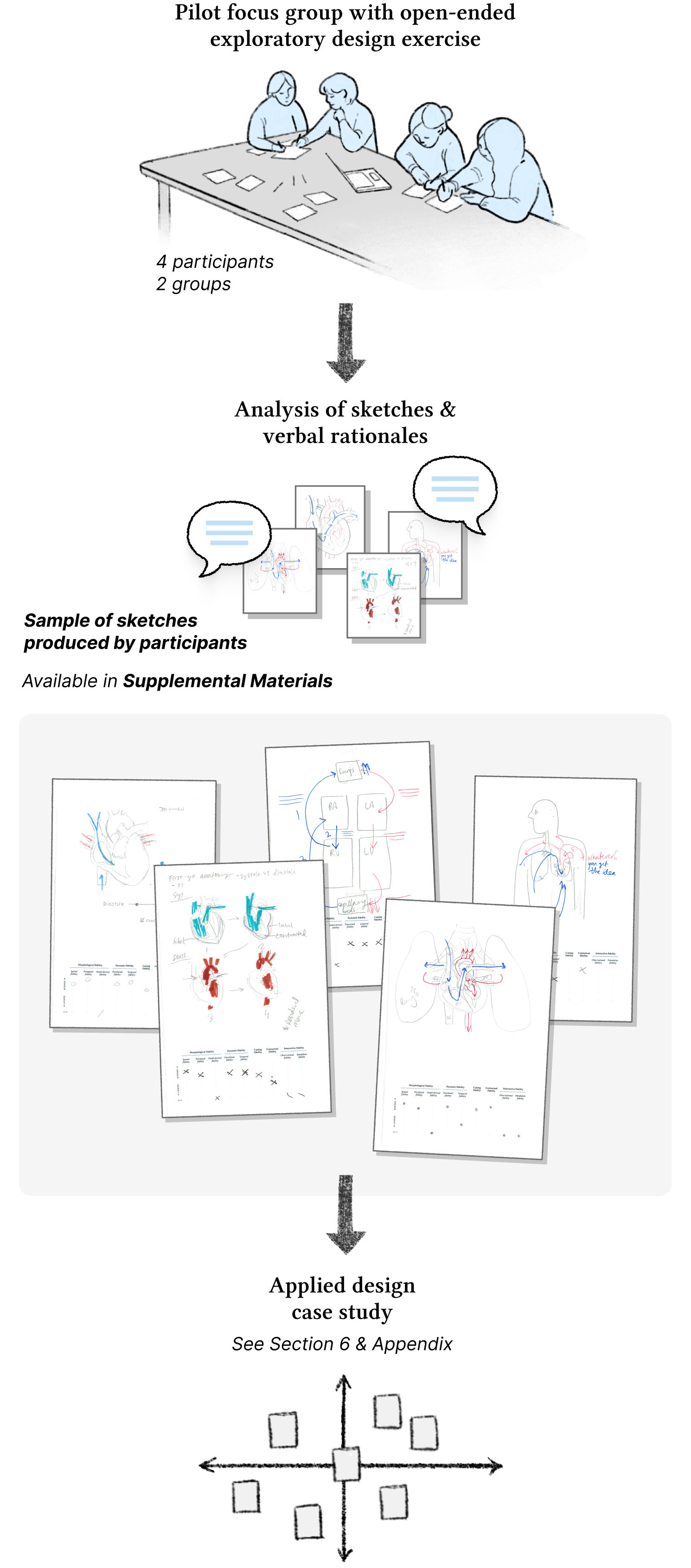}
  \caption{Methodological flow for Strand 2: from pilot focus group with open-ended exploratory design exercise, the analysis of sketches/verbal rationales, and finally how it informed the applied design case study presented below. We also share sample sketches produced by participants during the focus group. To see the complete set of sketches, and other relevant information, please see \hyperref[sec:Supplemental]{Supplemental Materials}, S3.}
  \label{fig:MethodStrand2}

\end{figure}

Figure~\ref{fig:MethodStrand2} illustrates the methodological flow for Strand 2 and includes sample sketches produced by participants during the focus group; for the complete set of sketches and other relevant materials, see \hyperref[sec:Supplemental]{Supplemental Materials}, S3.

\section{Design Space - Supplementary Guide}
\label{sec:appendixDisciminating}

This appendix provides supplementary guidance for each dimension and sub-dimension of the design space. For each, we provide the core discriminating question that anchors coding decisions, notes on boundary cases and common sources of ambiguity, and examples drawn from the corpus. This guide is intended to support researchers and practitioners in applying the design space consistently and reproducibly, and should be read alongside the dimension definitions in Section~\ref{sec:designspace} and the coded corpus in \hyperref[sec:Supplemental]{Supplemental Materials}, S1. A full record of how dimensions evolved across the three phases of corpus analysis (including dimensions that were proposed, revised, merged, or discarded) is available in \hyperref[sec:Supplemental]{Supplemental Materials}, S2.

\subsection{Morphological Fidelity}
Morphological Fidelity is the parent dimension encompassing three sub-dimensions (Spatial, Perceptual, and Model-derived Fidelity) that capture distinct aspects of how form and appearance are represented. The key organizing distinction concerns the nature of the referent and what kind of visual correspondence is possible. Spatial Fidelity applies when the visualization preserves structural and geometric relationships, regardless of surface appearance. Perceptual Fidelity applies when the referent has an observable appearance that the visualization approximates or reproduces. Model-derived Fidelity applies when the referent has no observable appearance and visual attributes are constructed rather than reproduced. A visualization may exhibit any combination of these three sub-dimensions simultaneously (e.g., a multi-panel figure that combines a rendering of an organ with a schematic of a related molecular process).

\subsubsection{Spatial Fidelity}
The core discriminating question for Spatial Fidelity is: does the visualization preserve the structural relationships of its referent (how parts are arranged, connected, and positioned relative to one another)? Spatial fidelity does not require photorealism or surface detail. A highly schematic line diagram can exhibit spatial fidelity if it preserves the structural geometry of its referent, for example, a topological diagram of the cardiovascular system that correctly represents which chambers connect to which vessels, even if proportions are stylized (see Figure~\ref{fig:DesignCaseStudy}). Figure~\ref{fig:CorpusExamples}C also illustrates this: a classical mechanics pulley diagram rendered entirely as geometric primitives and labeled lines, yet one that correctly preserves which masses connect to which pulleys via which ropes, and in what spatial configuration.

\subsubsection{Perceptual Fidelity}
The core discriminating question for Perceptual Fidelity is: does the referent have an observable appearance, and does the visualization reproduce or approximate that appearance through its visual attributes? This dimension applies only when the referent possesses an observable appearance, either directly (e.g., anatomical structures, everyday objects) or through extended human perception (e.g., microscopic imaging, where instruments amplify sensory access without constructing what is seen). Figure~\ref{fig:CorpusExamples}A illustrates this dimension clearly: a medical illustration of a laparoscopic splenectomy in which color, shading, tissue texture, and material properties correspond closely to how the anatomical structures would actually appear in a surgical context.

\subsubsection{Model-derived Fidelity}
The core discriminating question for Model-derived Fidelity is: does the visualization construct a visual appearance for an entity that has no observable appearance (which can be grounded in scientific data, computational models, disciplinary convention, etc.). This dimension uses a categorically different representational strategy appropriate to a different class of referent. A visualization of a protein folding pathway or a gravitational field cannot be evaluated against a perceptual standard because no such standard exists. Figure~\ref{fig:CorpusExamples}E illustrates this: a visualization of SARS-CoV-2 membrane fusion in which the molecular components are rendered with surface textures and material properties, a constructed and conventionally accepted visual proxy for entities that are directly imperceptible.

\subsection{Dynamic Fidelity}
Dynamic Fidelity is the parent dimension encompassing two sub-dimensions (Functional Fidelity and Temporal Fidelity) that capture distinct aspects of how change and behavior are represented. The key organizing distinction concerns what kind of dynamic information is encoded. Temporal Fidelity applies when the visualization conveys that the system evolves, transitions, or progresses through states over time. Functional Fidelity applies when the visualization encodes the causal and mechanistic relationships that drive that change. A visualization may exhibit either sub-dimension independently or both simultaneously. Temporal Fidelity without Functional Fidelity is common (e.g., an animation can show a process unfolding over time without conveying the mechanisms that drive it). Functional Fidelity without Temporal Fidelity is also possible, but less common based on our corpus examples, and subsequent analysis.

\subsubsection{Functional Fidelity}
The core discriminating question for Functional Fidelity is: does the visualization encode why or how the system behaves (i.e., the causal and mechanistic relationships that drive its behavior) or does it only show what the system looks like at one or more states? The presence of arrows does not automatically confer Functional Fidelity; for instance, arrows indicating a sequence without conveying causal mechanism may contribute to Temporal Fidelity but not Functional Fidelity. Functional Fidelity requires that the visualization encodes how or why one state produces the next, not merely that states differ. Figure~\ref{fig:CorpusExamples}F illustrates this: the interactive simulation encodes the mechanistic relationship between substrate concentration and reaction rate, allowing viewers to understand and predict causal outcomes (including the mathematical logic governing the relationship) rather than simply observe a visual state.

\subsubsection{Temporal Fidelity}
The core discriminating question for Temporal Fidelity is: does the visualization convey changes over time (e.g., that it evolves, transitions, or progresses through states) rather than simply showing what the system looks like at one moment or in multiple simultaneous configurations? A critical boundary applies: purely spatial motion (e.g., camera rotations, turntable animations, viewpoint changes) does not contribute to Temporal Fidelity unless the system itself changes in some way. A rotating 3D anatomical model may contribute to spatial fidelity by revealing structural detail from multiple viewpoints, but encodes no temporal change in the depicted system. Figure~\ref{fig:CorpusExamples}D showcases Temporal Fidelity: an animated infographic comparing respiratory mechanisms across humans, birds, and grasshoppers that encodes the cyclical temporal structure of each breathing process, including phase timing, directional flow, and the sequential progression of each respiratory cycle.

\subsection{Cueing Fidelity}
The core discriminating question for Cueing Fidelity is: are visual variables deployed to guide the viewer's attention toward specific information, relationships, or structural features? Figure~\ref{fig:CorpusExamples}B illustrates this clearly: a visualization of perivenous hepatic iron deposition in which the overall rendering is muted and naturalistic, while targeted color, callout lines, and annotation direct the viewer's attention to the specific hepatic structures and pathological features under discussion.

\subsection{Contextual Fidelity}
The core discriminating question for Contextual Fidelity is: does the visualization provide information that tells the viewer where or in relation to what the subject exists or operates? Context does not need to be realistic or complete to be present. Context can also emerge from relational structure among elements rather than from environmental background: when one element is clearly established as the primary subject and the remaining elements function as situating references, Contextual Fidelity is present even without an environmental backdrop. Figure~\ref{fig:CorpusExamples}G illustrates this: the Moveable Joint Types visualization situates the selected joint within the full skeletal structure. Panel C (classical mechanics pulley problem) provides a useful contrast: the mechanical system is presented against a neutral background without relational anchoring; it correctly represents structural relationships between components but provides no information about where the system exists or operates.

\subsection{Interactive Fidelity}
Interactive Fidelity is the parent dimension encompassing two sub-dimensions (Observational Fidelity and Simulation Fidelity) that capture a fundamental distinction in the nature of the interaction. The key organizing distinction concerns what the user's actions affect. Observational Fidelity applies when interactions alter how the system is seen without changing the system itself. Simulation Fidelity applies when interactions alter the system itself, producing consequential changes that reflect its causal, mechanistic, and/or mathematical logic. A visualization may exhibit either sub-dimension independently or both simultaneously. An important prerequisite applies to both: Interactive Fidelity is not achieved by the mere presence of interactive components. Interactions must expose structure, logic, relationships, and/or mechanism to qualify. A button that changes a background color does not contribute to Interactive Fidelity; a button that reveals internal anatomical structure, or that triggers a model-governed change in system behavior, does.

\subsubsection{Observational Fidelity}
The core discriminating question for Observational Fidelity is: can the user manipulate how the system is observed (i.e., changing their viewpoint, revealing hidden structure, or accessing information not available from a single fixed presentation) without changing the system itself? Typical interactions include rotating, zooming, panning, slicing, toggling layers, adjusting opacity, and switching representational modes. The defining criterion across all of these is that they alter how the system is seen, not what the system does. Figure~\ref{fig:CorpusExamples}G illustrates this: the Moveable Joint Types interactive visualization allows users to select different joint types and observe their movement.

\subsubsection{Simulation Fidelity}
The core discriminating question for Simulation Fidelity is: can the user manipulate parameters, conditions, or variables in ways that produce consequential changes within the represented system (i.e., changes that reflect the system's causal, mechanistic, or mathematical logic)? The defining criterion is that interactions modify the system, not merely the viewpoint, and that those modifications produce outcomes governed by the system's internal logic. The presence of sliders or input controls does not automatically confer Simulation Fidelity; the test is whether adjusting a control produces system-level changes. Figure~\ref{fig:CorpusExamples}F illustrates this: the substrate concentration simulation is governed by a mathematical model of enzyme kinetics, and adjusting substrate concentration produces model-computed changes in reaction rate displayed in real time.

\section{Corpus Examples}
\label{sec:appendixCorpusExamples}
Figure~\ref{fig:CorpusExamples} presents seven examples drawn from the corpus to illustrate the range of variation across all five primary dimensions and their sub-dimensions. Each example is annotated with its coded fidelity dimensions and discussed below to make the discriminating logic of each dimension concrete and legible. Readers are encouraged to consult the full coded corpus in \hyperref[sec:Supplemental]{Supplemental Materials}, S1, for a comprehensive view of how the dimensions are distributed and co-occur across the 175 items analyzed in this study.

\textbf{Figure~\ref{fig:CorpusExamples}A) Item 2 in Corpus — Division of Splenocolic Ligament.} This surgical illustration exhibits \textbf{Spatial Fidelity}, preserving the anatomical arrangement and relative positioning of the spleen, colon, and surrounding ligamentous structures. It exhibits \textbf{Perceptual Fidelity}: color, shading, tissue texture, and surface properties correspond closely to how these structures actually appear in a surgical field, satisfying the core discriminating question for this dimension (does the referent have an observable appearance, and does the visualization reproduce it?). The image also demonstrates \textbf{Contextual Fidelity}, situating the dissected structures within the broader anatomical field rather than isolating them against a neutral ground. \textbf{Cueing Fidelity} is present through the use of labels, line callouts, and symbols to direct the viewer's attention.

\textbf{Figure~\ref{fig:CorpusExamples}B) Item 97 in Corpus  — Perivenous Hepatic Iron Deposition in Alcoholic Cirrhosis.} This example is an anchor case for \textbf{Cueing Fidelity}: the overall rendering is deliberately muted and naturalistic, while targeted color, callout lines, and annotation direct attention to the specific hepatic structures and pathological features under discussion. It also exhibits \textbf{Spatial Fidelity} (the liver and surrounding vasculature preserve anatomical arrangement) and \textbf{Perceptual Fidelity} (tissue coloration and texture reflect observable appearance). \textbf{Contextual Fidelity} is present, as the affected liver is shown within a cross-sectional view of the abdominal cavity, establishing where the pathology occurs relative to surrounding organs. The visualization also exhibits \textbf{Model-derived Fidelity} in the callout depicting the magnetic field, an imperceptible phenomenon represented through a constructed visual convention. \textbf{Functional Fidelity} is present as well; for example, the visualization depicts the causal relationship between iron deposition and the resulting magnetic susceptibility property.

\textbf{Figure~\ref{fig:CorpusExamples}C) Item 134 in Corpus  — Classical Mechanics Problem (Pulleys).} This example depicts \textbf{Spatial Fidelity}: although rendered entirely as geometric primitives and labeled lines rather than realistic forms, it correctly preserves which masses connect to which pulleys via which ropes, and in what spatial configuration, satisfying Spatial Fidelity's discriminating question independent of surface realism. \textbf{Model-derived Fidelity} is present, as forces are represented using arrows, a constructed visual convention for an imperceptible quantity. \textbf{Functional Fidelity} is present, as the diagram encodes how the system behaves: the relationships between applied force, rope tension, and resulting mass are made explicit through the labeled variables and connections, allowing the mechanistic logic of the pulley system to be inferred. \textbf{Cueing Fidelity} is present through labeled variables (force vectors, lengths, masses) that direct attention to the quantities relevant to solving the problem. This example notably \textbf{lacks Contextual Fidelity}: the mechanical system is presented against a neutral background with no relational anchoring or indication of where the system exists or operates.

\textbf{Figure~\ref{fig:CorpusExamples}D) Item 9 in Corpus  — Three Different Ways to Breathe.} This animated infographic illustrates \textbf{Temporal Fidelity}; it encodes the cyclical temporal structure of human, bird, and grasshopper respiration, including phase timing, directional airflow, and the sequential progression of each respiratory cycle. It also exhibits \textbf{Functional Fidelity}, as the animation encodes how each respiratory system works, depicting the causal relationship between structural movement (e.g., diaphragm contraction, air sac expansion) and the resulting directional flow. It also exhibits \textbf{Spatial Fidelity}, preserving the relative arrangement of respiratory structures. \textbf{Contextual Fidelity} is present through the silhouetted body shapes/outlines that situate each respiratory system within its organismal context. \textbf{Cueing Fidelity} operates through color-coding and directional arrows that guide the viewer through the pathways of each species.

\textbf{Figure~\ref{fig:CorpusExamples}E) Item 17 in Corpus  — A Visual Model for SARS-CoV-2 Membrane Fusion.} This example depicts \textbf{Model-derived Fidelity}: the molecular components are rendered with surface textures and material properties that constitute a constructed, conventionally accepted visual proxy. \textbf{Spatial Fidelity} is present, preserving the structural arrangement of membrane components and viral proteins. \textbf{Temporal Fidelity} is present as the animation depicts the fusion process unfolding through sequential conformational states. Together, these dimensions support \textbf{Functional Fidelity}: the animation encodes how the underlying mechanisms work, conveying the causal relationships that drive membrane fusion, and its relationship to the molecular actors in the animation. \textbf{Cueing Fidelity} operates through selective coloration and emphasis at key moments in the animation, such as highlighting glycans situated atop the spike protein. Finally, \textbf{Contextual Fidelity} is present, as certain portions of the animation situate the virus relative to a larger host cell, anchoring the molecular-scale events within their broader biological setting.

\textbf{Figure~\ref{fig:CorpusExamples}F) Item 27 in Corpus  — Substrate Concentration (BIOMINT).} This interactive illustrates how \textbf{Functional}, \textbf{Observational}, and \textbf{Simulation Fidelity} work together: the simulation is governed by a mathematical model of enzyme kinetics, and adjusting substrate concentration produces visual changes in reaction rate displayed in an adjacent panel, with the system itself, not merely the viewpoint, changing in response to user input. The animation also conveys the nature of molecular movement and concentration over time, supporting \textbf{Temporal Fidelity}. \textbf{Spatial Fidelity} is present in the depiction of substrate-enzyme spatial relationships, while \textbf{Model-derived Fidelity} is present in the rendering of substrate molecules, an imperceptible referent represented through a constructed visual convention rather than reproduced appearance. \textbf{Contextual Fidelity} is also present: although the molecular interaction is shown without a literal environmental backdrop, the surrounding elements provide the situating and relational information needed to interpret the interaction's meaning. Finally, \textbf{Cueing Fidelity} is evident in the glow effect used to indicate molecular interaction.

\textbf{Figure~\ref{fig:CorpusExamples}G) Item 38 in Corpus  — Moveable Joint Types.} This interactive visualization is a good example of \textbf{Contextual Fidelity}: the selected joint is situated within the full skeletal structure. It also anchors \textbf{Observational Fidelity}: users can select different joint types and observe their movement, altering what is seen without altering the underlying system itself. \textbf{Spatial Fidelity} is present, preserving the structural arrangement of bones at each joint. \textbf{Perceptual Fidelity} is present, as the rendering of bone approximates observable anatomical appearance. \textbf{Cueing Fidelity} operates through highlighting what distinguishes the selected joint from the surrounding skeleton. \textbf{Functional Fidelity} and \textbf{Temporal Fidelity} are also present: selecting a joint reveals its characteristic range of motion unfolding over time, encoding both the mechanical relationship between connected bones and the cyclical or directional movement that joint type affords.

\begin{figure}[!th]
  \centering
  \includegraphics[width=0.46\textwidth, alt={(A) Bar charts showing the number of corpus items exhibiting each fidelity dimension and sub-dimension across 175 visualizations. (B) A heatmap matrix displaying Pearson correlation coefficients among dimensions and sub-dimensions, with color indicating positive or negative associations and gray cells indicating undefined correlations.}]{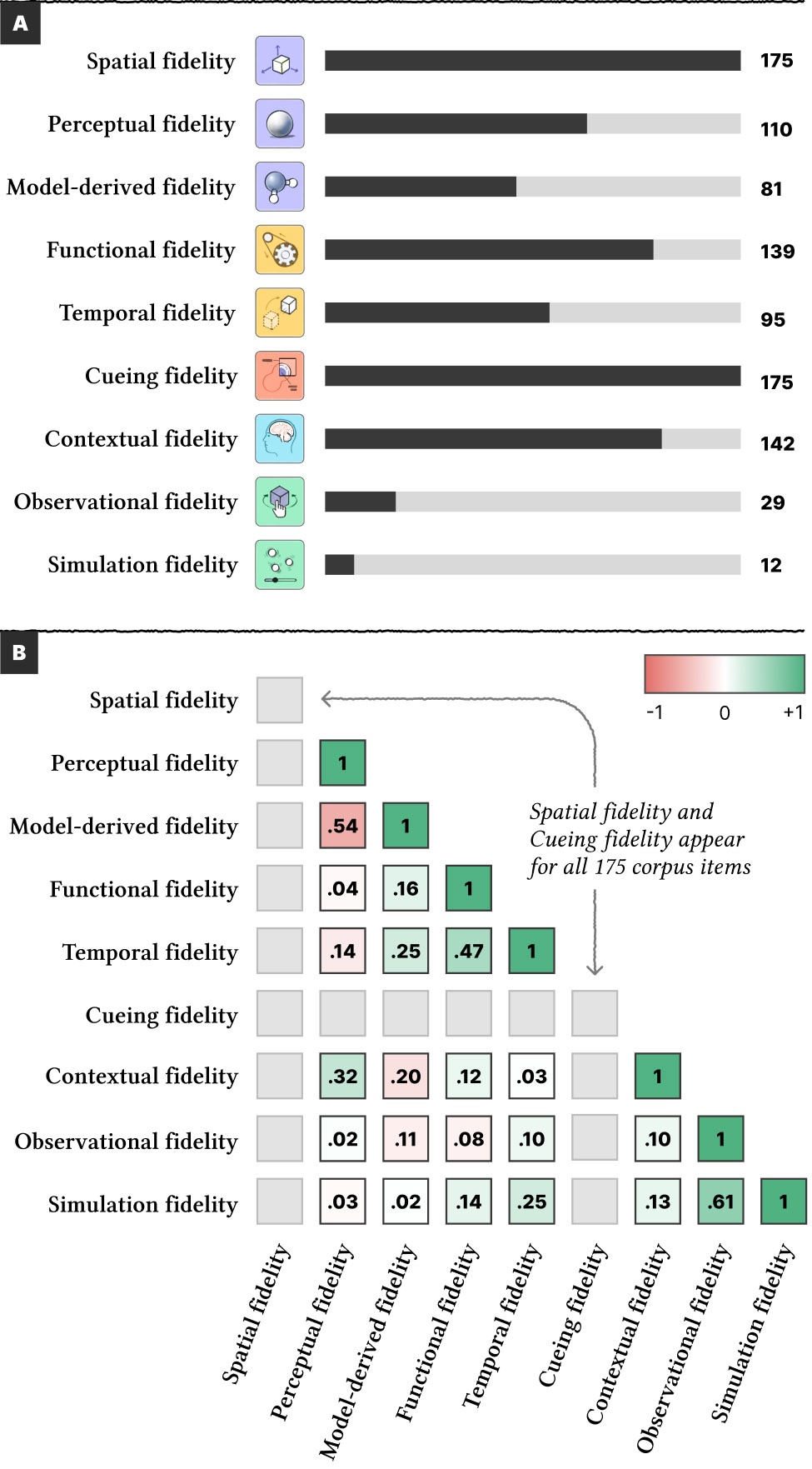}
  \caption{\textbf{(A)} Distribution of the incidence of all design dimensions and sub-dimensions across the final corpus. \textbf{(B)} Heat-map matrix showing Pearson correlation coefficients among all design dimensions and sub-dimensions. \textit{Spatial Fidelity} and \textit{Cueing Fidelity} are present in all 175 corpus items and therefore exhibit no variance; as a result, correlation coefficients involving these dimensions are not defined and are shown in gray. All remaining correlations are encoded using a diverging color palette, with green indicating positive correlations and red indicating negative correlations.}
  \label{fig:corpusanalysis}

\end{figure}

\begin{figure*}[!htbp]
  \centering
  \includegraphics[width=\textwidth, alt={Seven didactic visualizations are shown, labeled A–G, spanning different subject matter and media types. Examples include static medical illustrations, schematic engineering diagrams, animated explanatory graphics, and interactive visualizations. The visuals depict phenomena such as anatomical structures, mechanical systems, and biological processes, and vary in spatial structure, perceptual detail, depiction of system behavior over time, use of visual cueing, contextual framing, and interactivity. For each example, icons indicate which representational fidelity dimensions and sub-dimensions are present}]{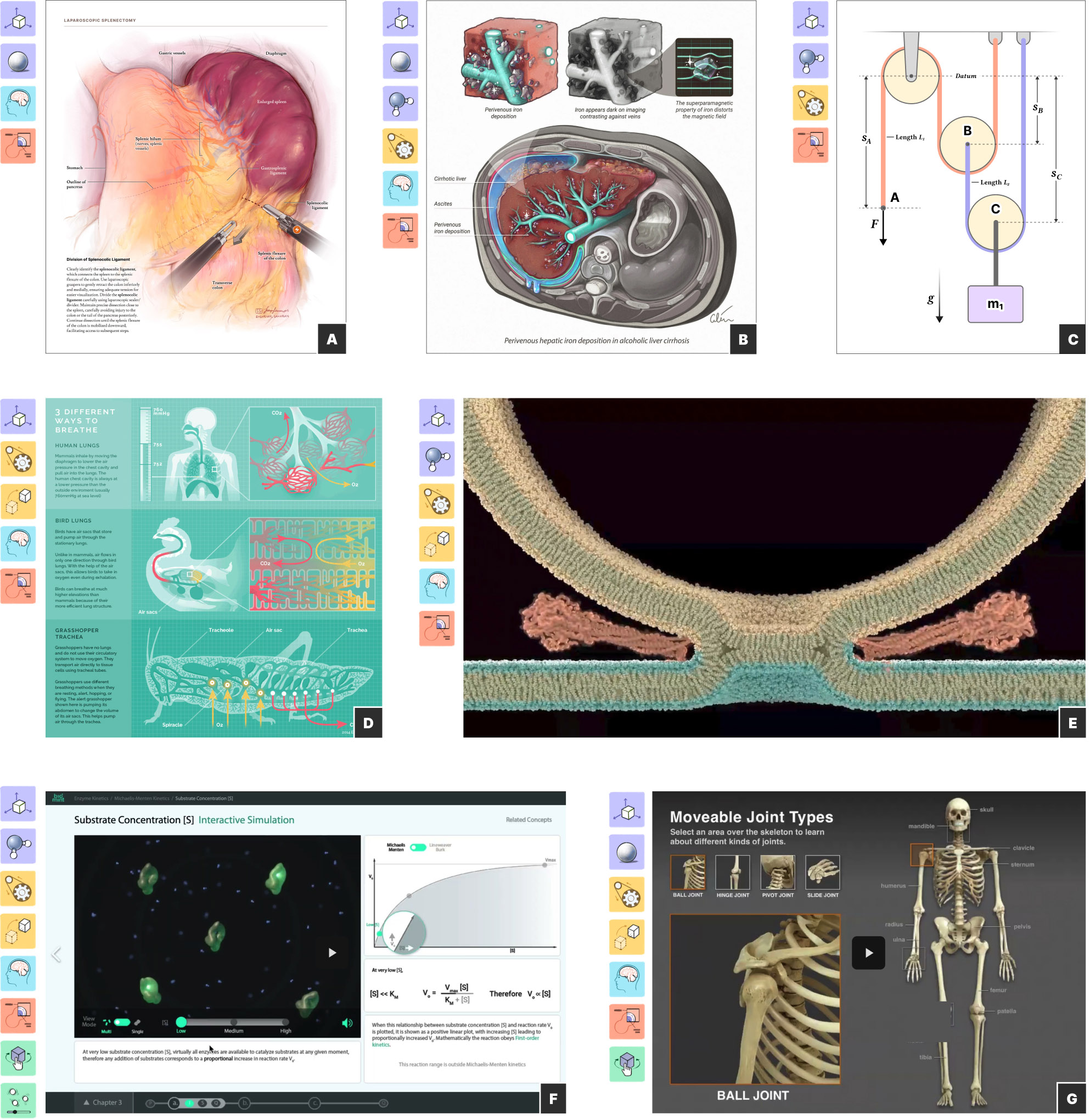}
  \caption{Corpus examples illustrating the design space of representational fidelity across subject matter and modalities. Seven visualizations are shown, arranged in three rows and labeled left to right, top to bottom (A–G). Each example is mapped to the proposed fidelity dimensions that best characterize its features. \textbf{A)} Item 2 — Division of Splenocolic Ligament — static visualization; Shehryar Saharan; \textbf{B)} Item 97 — Perivenous Hepatic Iron Deposition in Alcoholic Cirrhosis — static visualization; Artibiotics, Ciléin Kearns; \textbf{C)} Item 134 — Classical Mechanics Problem - Pulleys — static visualization; Shehryar Saharan; \textbf{D)} Item 9 — Three Different Ways to Breathe — animated looping visualization; Eleanor Lutz; \textbf{E)} Item 17 — A Visual Model for SARS-CoV-2 Membrane Fusion — animated visualization; Digizyme Inc.; \textbf{F)} Item 27 — Substrate Concentration, Biomolecular Interactive Tutorials (BIOMINT) — interactive visualization; Jerry Gu;\textbf{ G) }Item 38 — Moveable Joint Types (from E. O. Wilson’s \textit{Life on Earth} Unit 4, Interactive 16.1) — interactive visualization; Digizyme Inc.
}

  \label{fig:CorpusExamples}
\end{figure*}

\section{Corpus Analysis}
\label{sec:appendixCorrelation}
Figure~\ref{fig:corpusanalysis} presents the distributional and correlational analysis of the design space dimensions across the 175-item corpus. Panel A depicts the incidence of each dimension and sub-dimension across all corpus items. Panel B shows a heatmap matrix of Pearson correlation coefficients among all dimensions and sub-dimensions, computed over binary presence/absence coding. These data support the interpretation and discussion presented in Section \ref{sec:analysis}; readers are encouraged to consult the full coded corpus in \hyperref[sec:Supplemental]{Supplemental Materials}, S1, for item-level detail.

\section{Applied Design Case Study}
\label{sec:appendixAppliedDesignCase}
\begin{figure*}[!ht]
  \centering
  \includegraphics[width=0.93\linewidth, alt={A large composite figure mapping multiple design variants for the prompt “blood flow through the heart.” Starting from an initial baseline design, successive visualizations show systematic variation across fidelity dimensions, with icons beneath each image indicating the dimensions adjusted.}]{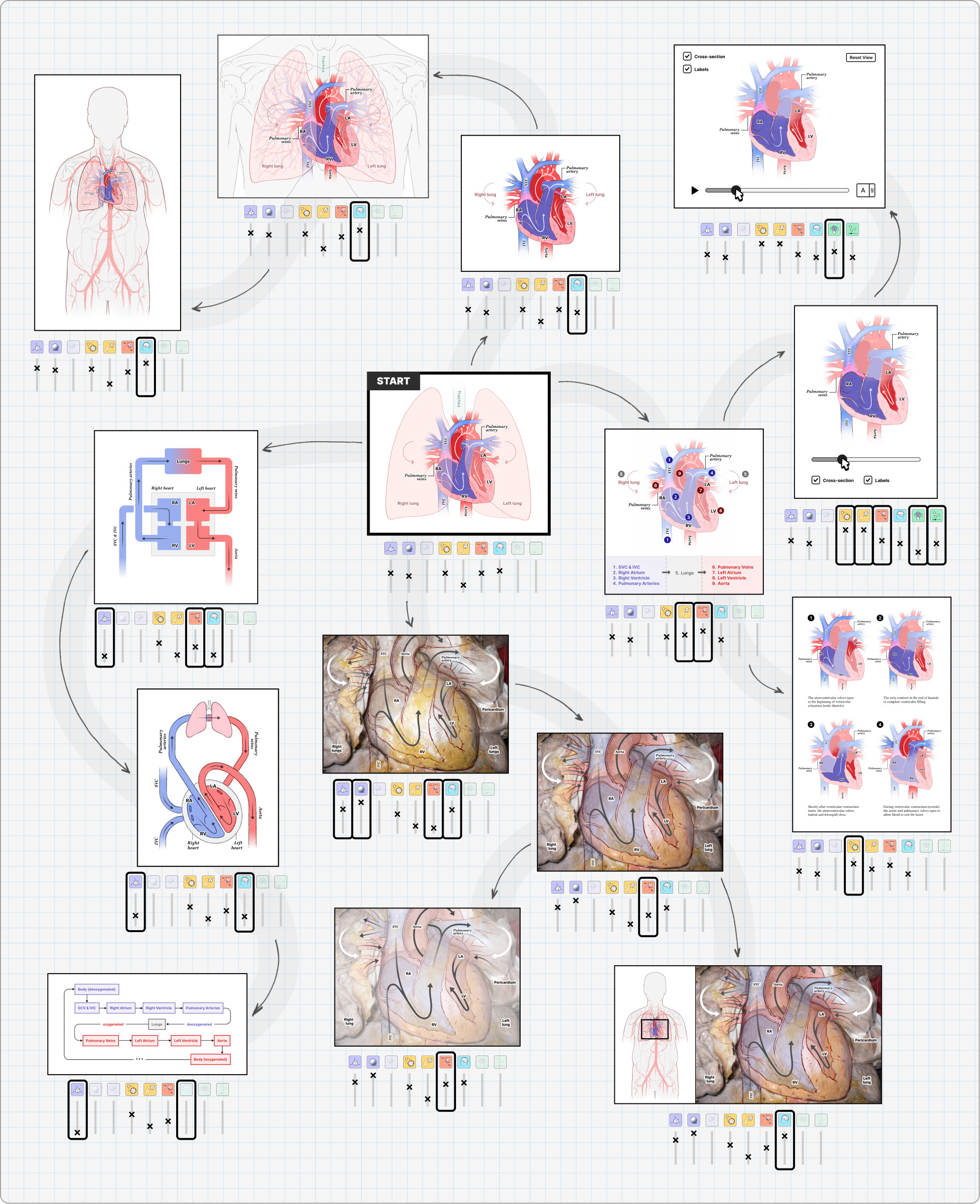}
   \caption{This figure maps an extended, self-directed traversal of visualization designs for the prompt, \textbf{\textit{blood flow through the heart}}, illustrating how the proposed design space is operationalized across multiple brainstorming and production sessions. Starting from an intuition-driven baseline (marked “START”), successive design variants reflect systematic manipulation of fidelity dimensions. Black squares beneath the variants indicate which dimensions of fidelity were altered relative to the previous iteration. Some visualizations in the lower half of the figure are based on a photograph of a heart dissection and have been illustrated over, annotated, and labeled [Heart dissection (28 October 2011), licensed under \href{https://creativecommons.org/licenses/by-sa/3.0/}{CC BY-SA 3.0}, via 
\href{https://commons.wikimedia.org/wiki/File:Heart_dissection.jpg}{Wikimedia Commons}]. All remaining illustrations were produced by the first author. A high-resolution version of this figure is available in \hyperref[sec:Supplemental]{Supplemental Materials}, S4.}

  \label{fig:DesignCaseStudy}

\end{figure*}

This appendix provides the full account of the applied design case study introduced in Section \ref{sec:strand1} and \ref{sec:applieddesignspacesection}, including annotated design variants (Figure~\ref{fig:DesignCaseStudy}), a walkthrough of the design process, and a discussion of the dimension indicators used throughout.
The dimension indicators shown beneath each variant are not reproducible measurement scores; they are reflective records of design intent, the first author's account of which dimensions were being foregrounded or suppressed during a given move. No consensus was sought among the author team, nor was any attempted, as this would misrepresent the purpose of the exercise. This strand is an exploration of the design space as a practitioner engages with it: traversing it through deliberate, situated decisions rather than measuring fixed positions within it. Reproducible coding criteria are the appropriate standard for Strand 1 (Section \ref{sec:methods}), where binary presence/absence coding was used and consensus was reached; they are not the appropriate standard here, where individual judgment and creative risk-taking are the point. 

This also bears directly on how Figure \ref{fig:HybridSpace} should be interpreted. The position of each design variant along the Spatial and Perceptual Fidelity axes reflects the first author's holistic judgment, consistent with how these dimensions were used throughout this strand: as a communicative and exploratory tool for design reasoning, not as a precise or reproducible measurement. Strand 1's binary coding (Section \ref{sec:strand1}) served that role for the corpus. A more substantive open question, separate from precision of placement, is this: within a single dimension, fidelity can vary across different regions of the same visualization, for example, a relatively low Perceptual-Fidelity contextual illustration (e.g., a simplified body silhouette establishing anatomical location) positioned adjacent to a larger, higher Perceptual-Fidelity focal image (e.g., a photorealistic rendering of the organ under discussion). Our binary, presence/absence coding does not capture this within-dimension variation; a visualization is coded as exhibiting Perceptual Fidelity if it appears anywhere within the image, regardless of how much of the image it characterizes. We do not attempt to resolve how such variation should be reconciled into a single characterization of the dimension in this particular work.

\textbf{Walkthrough of the Applied Design Case Study:} To further clarify how the applied design case study unfolded in practice, the first author briefly reflects on the design process that produced the variants mapped in Figure~\ref{fig:DesignCaseStudy}:

\vspace{0.5em}
\begin{adjustwidth}{1.2em}{0em}
 \textit{
 I began with a didactic visualization grounded primarily in intuitive design judgment, mirroring the patterns observed in the pilot focus group. The initial design exhibited a particular balance of cueing, spatial structure, perceptual detail, and contextual information. From this baseline, I transitioned into a structured ideation phase that explicitly engaged with the proposed design space. During this phase, the fidelity dimensions were actively manipulated as generative dials (dimensions are shown below each sketch in Figure~\ref{fig:DesignCaseStudy}), supporting deliberate exploration and partial saturation of the space.
Contextual Fidelity proved to be a productive early lever. I began by decreasing contextual information to simplify the visual, then progressively reintroduced it, first providing the required anatomical situation, then adding secondary structures that were not the primary instructional focus but served a supplementary role, and finally extending context further to provide a more familiar, whole-body frame of reference. From there, I explored the interaction between perceptual and spatial fidelity, which proved less predictable than I expected. Stripping Perceptual Fidelity entirely while preserving spatial relationships produced clean, legible variants, but reintroducing perceptual detail incrementally exposed an interesting middle ground between faithful depiction and schematized rendering. I then pushed spatial fidelity to its lowest possible level, generating the most abstracted configurations I could imagine. Rather than converging on a single "best" option, these variants prompted me to consider audience: not about who the audience is (...that was already defined) but about when and why a given configuration would serve them. At what stage of learning would this level of abstraction be beneficial? Which learning goals would it support, and which might it undermine? Could the same variant that scaffolds early conceptual understanding become a hindrance once a learner needs to reason about structural detail? Later moves explored how Cueing Fidelity could be layered alongside other dimensions to foreground temporal information. I experimented with different formats for encoding sequence and change, and considered how these might influence Observational or Simulation Fidelity. Each of these design moves is captured in Figure~\ref{fig:DesignCaseStudy}.}
\end{adjustwidth}
\vspace{1em}

% \textbf{Reflections:} Based on this early applied work, several tentative observations emerge. The fidelity dimensions appear to function generatively, enabling a form of systematic exploration that extends beyond intuition or convention alone, though the extent to which this holds across different designers, prompts, and domains remains an open question. Dimensions also seem to interact in ways that produce emergent effects: adjustments along one dimension frequently reshaped the communicative role of others, suggesting that the representational impact of any single dimension cannot be fully anticipated in isolation. We are cautious about generalizing from a single practitioner's traversal, and acknowledge that a different designer working from the same prompt would likely produce a different set of variants and encounter different points of friction. That variability, however, may itself be informative: it would test whether the design space consistently supports structured exploration even when the specific paths through it differ. Together, these observations offer preliminary (but we believe encouraging) support for the generative and reflective claims made in this work, while also underscoring the need for broader engagement with the framework across a wider range of practitioners, audiences, and instructional contexts.

\end{document}